\documentclass{emulateapj}
\usepackage{amsmath}  
\usepackage{natbib}  
\usepackage[export]{adjustbox}
\usepackage[dvipsnames]{xcolor}  %% colors
\usepackage[colorlinks=true,
            breaklinks=true,
            linkcolor=NavyBlue,
            citecolor=NavyBlue,
            urlcolor=NavyBlue]{hyperref}
\usepackage{booktabs} %% table cmidrule

\usepackage{enumitem}
\setlist[enumerate]{itemindent=\dimexpr\labelwidth+\labelsep\relax,leftmargin=0pt}

\usepackage[normalem]{ulem} %% strikethrough line

\newcommand{\Chandra}[0]{\textit{Chandra}}
\newcommand{\chandra}[0]{\textit{Chandra}}

\newcommand\ergs{erg~s$^{-1}$}

\newcommand{\solarmass}{$M_{\sun}$}
\newcommand{\sfr}{M$_{\sun}$~yr$^{-1}$}

\newcommand{\mbh}{$M_{\rm BH}$}
\newcommand{\msigma}{$M_{\rm BH}-\sigma$}

\newcommand{\eddr}{$\lambda_{\rm Edd}$}

\newcommand{\nhint}{$N_{\rm H}$}

\newcommand{\lx}{$L_{\rm X}$}
\newcommand{\lha}{$L_{\rm H\alpha}$}
\newcommand{\lbol}{$L_{\rm bol}$}
\newcommand{\ledd}{$L_{\rm Edd}$}

\newcommand{\halpha}{H{$\alpha$}}

\newcommand{\OI}{[O\,{\sc i}]}

\newcommand{\NII}{[N\,{\sc ii}]}

\newcommand{\SII}{[S\,{\sc ii}]}

\newcommand{\HII}{H\,{\sc ii}}

\newcommand{\GalNum}{719}        %% Galaxy Number
\newcommand{\AGNNum}{314}           %% AGN Number
\newcommand{\MbhNum}{250}           %% AGN with Mbh
\newcommand{\onesec}{228}          %% Number of AGN within 1 arcsec
\newcommand{\AGNFrac}{43.7}       %% AGN Fraction
\newcommand{\BarNum}{285}        %% Galaxy Number with BAR

\newcommand{\ClassNum}{418}         %% number of classification
\newcommand{\ClassHoNum}{249}       %% number of classification from Ho
\newcommand{\ClassVeronNum}{44}     %% number of classification from Veron
\newcommand{\AClassNum}{243}        %% number of AGNs have classification

\newcommand{\SeyNum}{59}
\newcommand{\LinNum}{66}
\newcommand{\TraNum}{41}
\newcommand{\HiiNum}{163}
\newcommand{\AbpNum}{89}
\begin{document}
\title{Chandra Survey of Nearby Galaxies: A Significant Population of Candidate Central Black Holes in Late-type Galaxies}
\shorttitle{Central Black Holes in Late-type Galaxies}
\shortauthors{She, Ho, \& Feng}

\author{Rui She\altaffilmark{1},
Luis C. Ho\altaffilmark{2,3}, and
Hua Feng\altaffilmark{1}}

\altaffiltext{1}{Department of Engineering Physics and Center for Astrophysics, Tsinghua University, Beijing 100084, China}
\altaffiltext{2}{Kavli Institute for Astronomy and Astrophysics, Peking University, Beijing 100087, China}
\altaffiltext{3}{Department of Astronomy, School of Physics, Peking University, Beijing 100087, China}
%\email{sher12@mails.tsinghua.edu.cn}

\tabletypesize{\scriptsize}

\begin{abstract}
  Based on the \chandra\ data archive as of March 2016, we have identified \AGNNum\ candidate active galactic nuclei in \GalNum\ galaxies located closer than 50 Mpc, among them late-type (Hubble types Sc and later) galaxies that previously had been classified from optical observations as containing star-forming (\HII) nuclei.  These late-type galaxies comprise a valuable subsample to search for low-mass ($\lesssim 10^6$~\solarmass) central black holes.  For the sample as a whole, the overall dependence of the fraction of active nuclei on galaxy type and nuclear spectral classification is consistent with previous results based on optical surveys.  We detect 51 X-ray cores among the \HiiNum\ \HII\ nuclei and estimate that, very conservatively, $\sim$74\% of them with luminosities above $10^{38}$~\ergs\ are not contaminated by X-ray binaries; the fraction increases to $\sim$92\% for X-ray cores with a luminosity of $10^{39}$~\ergs\ or higher. This allows us to estimate a black hole occupation fraction of $\gtrsim 21$\% in these late-type, many bulgeless, galaxies.
\end{abstract}


%% introduction part 
\section{Introduction}
\label{sec:intro}

Nearby galaxies offer us the unique laboratories to search for and study weak active galactic nuclei (AGNs), including those emanating from low-mass central black holes ($M_{\rm BH} < 10^6$~\solarmass).  Although their formation mechanism is still poorly constrained, low-mass black holes may have played a key role in forming supermassive black holes (SMBHs) in the early universe.  The early manifestation of SMBHs at high redshifts \citep{mortlock11,wu15} requires the growth of these giants to start with ``seed'' black holes that are significantly more massive than stellar-mass black holes of typical mass 10~\solarmass \citep{volonteri10}.  Some of the seed black holes may have survived until today and appear as low-mass black holes lurking at the centers of low-mass (late-type spiral and dwarf) galaxies \citep{volonteri08,greene12,reines16a}.  Present-day low-mass galaxies, to the extent that they have remained relatively unevolved since their formation, are the optimal sites to search for such relic seed black holes.  Finding these low-mass black holes, knowing their occupation fraction in low-mass galaxies, and measuring their mass function may help constrain their formation mechanism and distinguish between the two competing models: whether they are Population III remnants \citep{heger03} or due to direct collapse \citep{haehnelt93}. 

According to the scaling relations between black hole mass and host galaxy properties \citep{kormendy13}, the best place to unveil low-mass black holes is in late-type bulgeless galaxies. Optical and X-ray searches have been conducted in the past. Systematic searches for low-mass black holes using the Sloan Digital Sky Survey uncovered several hundred new objects residing in low-mass galaxies \citep{greene04,greene07b,dong12,reines13}, but the statistical completeness of these optical searches are difficult to quantify \citep{greene09}, as AGN samples selected by optical emission lines tend to have black holes radiating at moderate to high fractions of their Eddington limits. On the contrary, X-ray observations, especially of high resolution and high sensitivity afforded by the \chandra\ Advanced CCD Imaging Spectrometer \citep[ACIS;][]{weisskopf02},  can reveal very weak nuclear activity in galaxies, even with brief exposures \citep[$\sim$ks; e.g.,][]{ho01,gallo08,miller12}. 
\citet{desroches09a} performed the first dedicated search for low-mass black holes with \chandra\ and detected 17 candidates from 64 late-type spirals.  Later on, cross-matching nearby dwarf galaxies with \chandra\ and {\it XMM-Newton} catalogs of X-ray point-like sources increased the sample size to $\sim$93 \citep{lemons15,pardo16,nucita17}. Using stacked \chandra\ images, \citet{mezcua16} revealed a population of low-mass black holes that are too faint to be detected individually in dwarf galaxies. The hard X-ray {\it NuSTAR}\ observations of \citet{chen17}, albeit of a small sample, also suggested a non-negligible fraction of accreting low-mass central black holes that may have been missed in optical surveys.

% introduce Paper I
Here we report the search of low-mass black holes from an even larger sample of nearby late-type galaxies.
We have recently conducted an X-ray survey of a large sample of nearby galaxies. The main goals of this survey, described in \citet[][hereafter Paper I]{she17}, are to systematically search for and study AGNs  in all galaxies within 50 Mpc contained in the \Chandra\ archives observed with ACIS. The sample contains \GalNum\ galaxies, of which \AGNNum\  are identified as AGN candidates by the cross-correlation of X-ray point-like sources and the near-infrared or optical stellar nuclei of these galaxies. The technical details of the survey and the X-ray properties of the AGN candidates are presented in Paper I. With 2$-$10 keV luminosities less than 10$^{42}$~\ergs\ for most objects, these sources occupy the regime of low-luminosity AGNs \citep[for a review, see][]{ho08}

% This paper focus
This paper focuses on the identification of a population of previously hidden AGNs in the centers of late-type, mostly bulgeless galaxies previously classified as \HII\ nuclei in optical surveys. Most of them have very low luminosities and are mostly likely associated with low-mass black holes.  Unlikely their optically identified counterparts \citep[e.g.,][]{greene04,greene07b,dong12,reines13}, this X-ray--selected population, having significantly lower accretion rates, is much more numerous, allowing us to establish a new, firmer estimate of the occupation fraction of low-mass black holes in the nearby universe.   For completeness, the paper also addresses the overall demographics of AGNs in the full sample of nearby galaxies, which covers a broad range of Hubble types. 

Section~\ref{sec:data} gives a brief introduction of the sample and measurements. Section \ref{sec:demography} presents the detection rates of X-ray AGN candidates in different kinds of galaxies and their possible dependence on the presence of a large-scale bar. Black hole masses and Eddington ratios are estimated in Section~\ref{sec:mbh_edd}. Section \ref{sec:hii} highlights the properties of X-ray AGN candidates associated with \HII\ nuclei. We summarize our results in Section~\ref{sec:conclusion}.

\section{Sample and Data Analysis}
\label{sec:data}

Paper I gives a detailed description of the sample and data reduction.  Here we just give a brief summary of the most important aspects.
The sample is assembled based on the full list of \Chandra\ imaging observations, with both ACIS-S and ACIS-I, as of March 2016.
All \chandra-observed galaxies with a distance of less than 50~Mpc were included in our sample, resulting in a total of \GalNum\ galaxies.

%% --- background, source, definition of Hubble type, bar, classifications ----
The Hubble types of the galaxies come from the NASA/IPAC Extragalactic Database (NED),\footnote{{\tt http://ned.ipac.caltech.edu}} which are taken largely from the Third Reference Catalogue of Bright Galaxies (RC3; \citealt{devaucouleurs91}). A total of \BarNum\ galaxies have a large-scale bar structure (with Hubble type SAB or SB).
Optical nuclear spectral classifications are available for \ClassHoNum\ galaxies from the Palomar spectroscopic survey of bright, northern galaxies \citep{ho95,ho97}. Nuclear spectral classifications for another \ClassVeronNum\  galaxies are found in the catalog of \citet{veron-cetty10}. For the 125 galaxies whose spectral classifications are not included in these two catalogs, we performed our own classification using optical emission-line ratios. Methods of spectral fitting and classification can be found in Paper I. To summarize:  \ClassNum\ out of \GalNum\ galaxies in our sample have an optical nuclear spectral classification \citep[for a review see][]{ho08}, including \SeyNum\ Seyferts, \LinNum\ low-ionization nuclear emission-line regions \citep[LINERs;][]{heckman80a}, \TraNum\ transition objects (emission-line nuclei with \OI\ strengths intermediate between those of \HII\ nuclei and LINERs; see \citealt{ho93}), \HiiNum\ \HII\ nuclei, and \AbpNum\ absorption-line nuclei (those without emission lines). 

Point-like X-ray sources are detected using CIAO task \textit{wavdetect}, and then AGN candidates are identified by detection of an X-ray core astrometrically coincident with the near-infrared/optical position of the galaxy nucleus, taking various errors into account. This results in \AGNNum\ X-ray AGN candidates, of which \onesec\ are located in the sky plane less than 1$\arcsec$ from the near-infrared/optical nucleus. X-ray spectral fitting is conducted for AGN candidates with enough photons; otherwise, hardness ratios are obtained to characterize the shape of the spectrum. The X-ray luminosity in the 2$-$10 keV band of each AGN candidate is derived from spectral fitting if available, or from count rates otherwise. For \MbhNum\ objects, Eddington ratios are calculated based on black holes masses derived from the \msigma\ relation\footnote{The \msigma\ relation we used in Paper I is $\log [M_{\rm BH}/(10^9~M_{\sun})]=-(0.68\pm0.05)+(5.20\pm0.37)\log[\sigma/(200~{\rm km~s^{-1}})]$.} \citep[][as supplemented in Paper I]{kormendy13}.

\section{Demographics of AGNs}
\label{sec:demography}

\begin{figure}[t]
\centering
    \includegraphics[width=0.7\columnwidth]{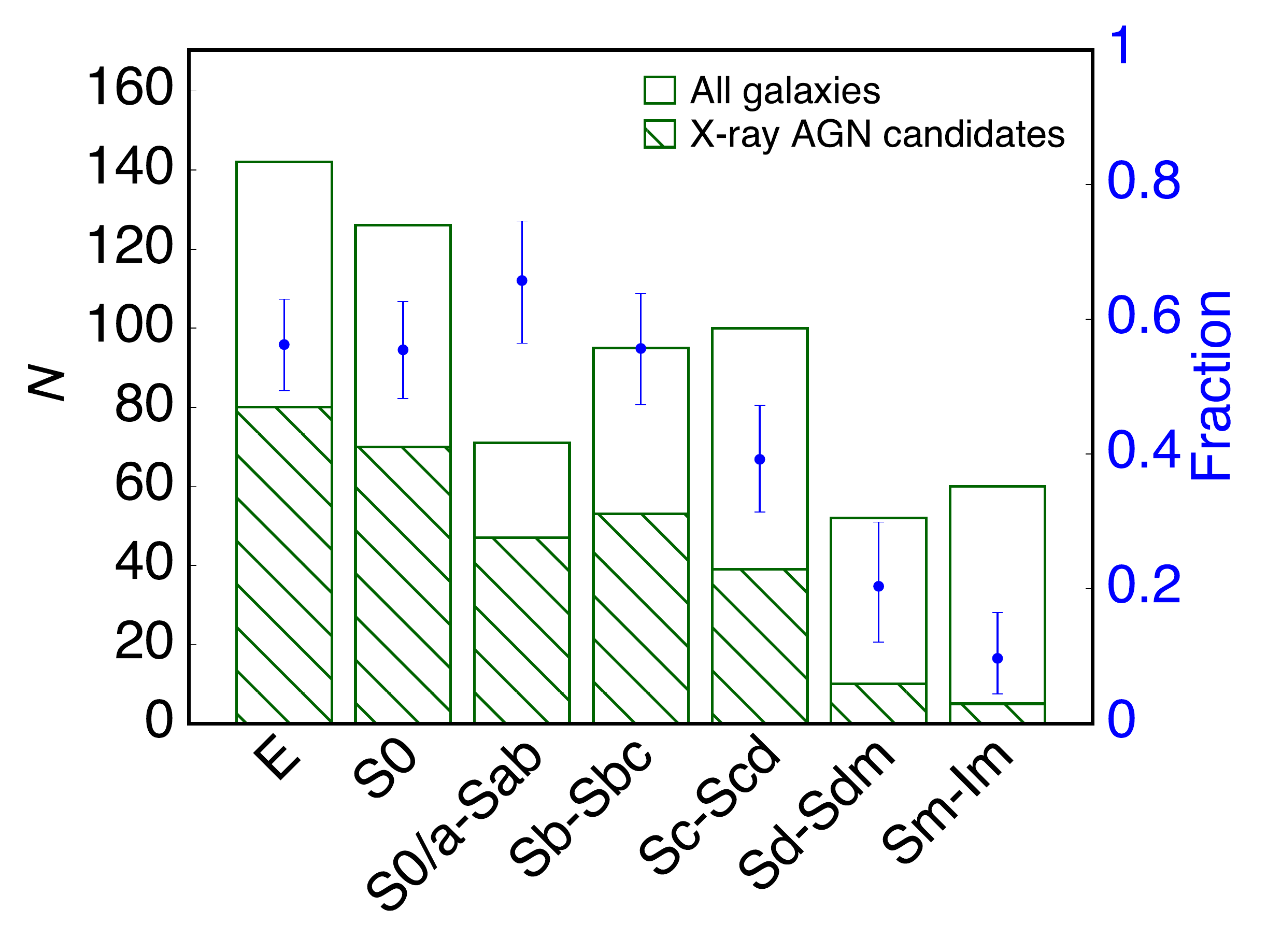}
    \includegraphics[width=0.7\columnwidth]{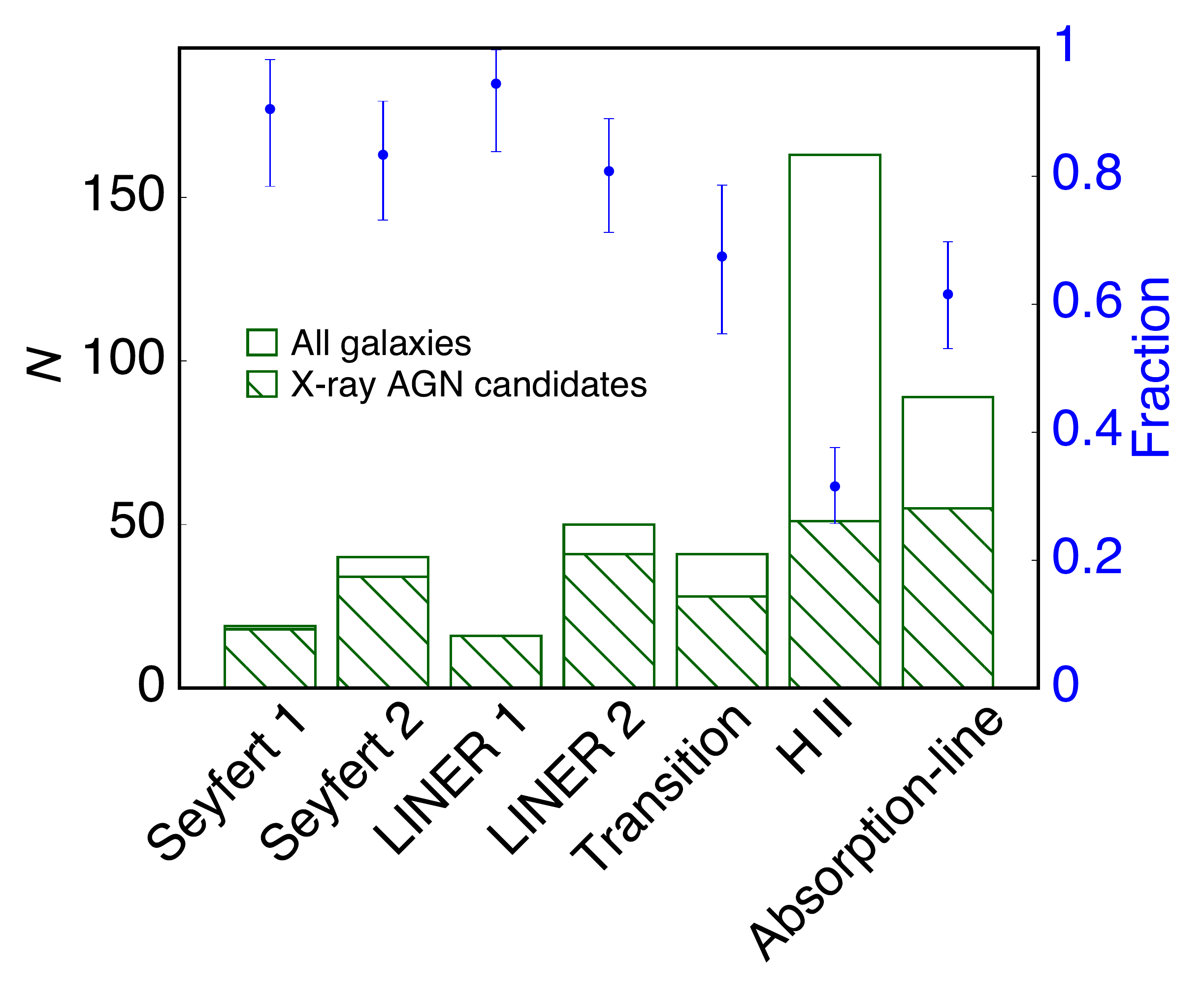}
  \caption{Number of galaxies (open histograms), number of X-ray AGNs (hatched histograms), and the fraction of AGNs (blue dots with 90\% errors) as a function of the galaxy Hubble type (left) and optical nuclear spectral classification (right).
  \label{fig:coincident}}
\end{figure}

%% ------------ AGNs in types ------------ 
The numbers of galaxies and AGN candidates as a function of Hubble type are shown in Table~\ref{tab:agnrates} and in Figure~\ref{fig:coincident} (left), which also displays the X-ray core detection rate with a 90\% error range for each type. For early-type galaxies (E$-$Sbc), the X-ray core detection rate is as high as $\sim$60\% (250/434), whereas for  late-type galaxies (Sc$-$Im), the X-ray core detection rate drops to $\sim$25\% (54/212). For galaxies with a Hubble type of Sm$-$Im, at the extreme end of the late-type sequence, the fraction is less than 10\%. The X-ray core detection rate in all galaxies is \AGNFrac\%. These AGN detection rates are broadly consistent with those derived from optical spectroscopic observations \citep{ho97b,ho08}, which reported AGN fractions of 50\%$-$70\% in early-type galaxies and $\sim$15\% in late-type systems. There appears to be a slight overabundance of X-ray AGN candidates among Sc$-$Sdm galaxies (see Table~\ref{tab:agnrates}) compared to optical AGN detection rate found in the Palomar survey. X-ray observations are more sensitive to weak AGNs than optical observations. The results in this work largely confirm the findings of former X-ray investigations  \citep{ho08,ho09,desroches09a,zhang09,grier11}.  Compared to previous similar X-ray studies, this work targets a much larger sample of galaxies and utilizes high-quality, high-resolution \chandra\ data, which were uniformly analyzed.  We also perform a  more comprehensive analysis on position errors, which is critical to evaluate whether any given compact X-ray source is likely to be associated with the nucleus of the galaxy.

%%%% -------- Table of Hubble types and Bar -------------
%\floattable 
\begin{deluxetable*}{lrrcrrcrrr}
\tabletypesize{\scriptsize}
\tablewidth{0pt}
\tablecaption{AGN numbers and detection rates as a function of Hubble type}
\tablehead{
\colhead{Type}                       &
\multicolumn{3}{c}{All Galaxies } &
\multicolumn{3}{c}{Unbarred Galaxies} &
\multicolumn{3}{c}{Barred Galaxies}  \\
\cmidrule(lr){2-4} 
\cmidrule(lr){5-7} 
\cmidrule(lr){8-10} 
                                     & 
\colhead{$N_{\rm G}$}                      &
\colhead{$N_{\rm A}$}                      &
\colhead{Rate (\%)}                   &
\colhead{$N_{\rm G}$}                      &
\colhead{$N_{\rm A}$}                      &
\colhead{Rate (\%)}                   &
\colhead{$N_{\rm G}$}                      &
\colhead{$N_{\rm A}$}                      &
\colhead{Rate (\%)}                   }
\startdata
 E        & 142 & 80  & 56$^{+6}_{-6}$   & \nodata & \nodata & \nodata          & \nodata & \nodata & \nodata          \\
 S0       & 126 & 70  & 55$^{+7}_{-7}$   & 37      & 26      & 69$^{+11}_{-12}$ & 51      & 27      & 52$^{+11}_{-11}$ \\
 S0/a-Sab & 71  & 47  & 65$^{+8}_{-9}$   & 22      & 17      & 75$^{+12}_{-15}$ & 38      & 24      & 62$^{+12}_{-12}$ \\
 Sb-Sbc   & 95  & 53  & 55$^{+8}_{-8}$   & 29      & 16      & 54$^{+14}_{-14}$ & 60      & 35      & 58$^{+10}_{-10}$ \\
 Sc-Scd   & 100 & 39  & 39$^{+8}_{-7}$   & 34      & 10      & 30$^{+13}_{-11}$ & 58      & 29      & 50$^{+10}_{-10}$ \\
 Sd-Sdm   & 52  & 10  & 20$^{+9}_{-8}$   & 11      & 4       & 38$^{+22}_{-20}$ & 37      & 6       & 17$^{+10}_{-8}$  \\
 Sm-Im    & 60  & 5   & 9$^{+6}_{-5}$    & 4       & 2       & 50$^{+31}_{-31}$ & 40      & 2       & 7$^{+7}_{-5}$    \\
 I0       & 6   & 4   & 62$^{+24}_{-28}$ & \nodata & \nodata & \nodata          & \nodata & \nodata & \nodata          \\
 pec      & 6   & 1   & 25$^{+27}_{-19}$ & \nodata & \nodata & \nodata          & \nodata & \nodata & \nodata          \\
 Unknown  & 61  & 5   & 9$^{+6}_{-5}$    & \nodata & \nodata & \nodata          & \nodata & \nodata & \nodata          \\
 All      & 719 & 314 & 43$^{+3}_{-3}$   & 137     & 75      & 54$^{+6}_{-6}$   & 284     & 123     & 43$^{+4}_{-4}$   \\

\enddata
\label{tab:agnrates}
\tablecomments{The number of galaxies and AGN candidates in different bins of Hubble type, for all galaxies, galaxies without a bar (SA), and galaxies with a bar (SAB and SB). $N_{\rm G}$ is the number of galaxies, and $N_{\rm A}$ is the number of AGN candidates. For AGN detection rates, the errors are given at a 90\% confidence level.}% with a Bayesian approach.} 
\end{deluxetable*}

%%% -------  CLASSES

%% ------- AGN in class --------
Out of the \ClassNum\ galaxies with reliable optical spectral classifications, \AClassNum\ (58\%) have an X-ray AGN candidate. The X-ray AGN detection rate corresponding to each optical spectral classification is shown in Table~\ref{tab:spectral_class} and in Figure~\ref{fig:coincident} (right). The frequency of X-ray AGNs is high in optical-selected Seyferts and LINERs, $>$90\% for type 1 (broad-line) and $\sim$80\% for type 2 (narrow-line) sources. About 70\% of the transition objects contain an X-ray nucleus. Pure \HII\ nuclei have the lowest detection rate of X-ray counterparts, but still $>30$\%. Approximately 60\% of absorption-line nuclei harbor an X-ray AGN candidate. These results are broadly consistent with those in \citet{ho08} and \citet{zhang09}.
For all X-ray AGN candidates with an optical spectral classification, 21\% are Seyferts (7\% type 1 and 14\% type 2), 24\% are LINERs (7\% type 1 and 17\% type 2), 12\% are transition objects, 21\% are \HII\ nuclei, and 22\% are absorption-line nuclei. As a result, more than 40\% of X-ray AGN candidates (\HII\ and absorption-line nuclei) in this work were previously completely missed in optical surveys.

%%% --- notes on absorption-line nuclei
It is worth noting that, most of absorption-line nuclei are early-type galaxies (77/89 are ellipticals and lenticulars), all of which should harbor a central black hole \citep{kormendy13}. The dearth of cool gas in these galaxies presumably leads to very low mass accretion rates for their central black holes, resulting in very low accretion-powered luminosities and consequently undetected optical line emission, especially in the presence of strong contamination by host galaxy starlight.
%%% -- non-detection in Seyferts and LINERs
For the 16 Seyferts and LINERs without a detection of an X-ray core, the most likely culprit is sensitivity. The exposure time is less than 6 ks for most of them (10/16), and for 13 of them the 3$\sigma$ upper limit of the X-ray luminosity in 2$-$10 keV is higher than $10^{38}$~\ergs. For those having \halpha\ measurements, the 10$-$90 percentile range of their X-ray (upper limit) to \halpha\ luminosity ratios is 0.2$-$10, consistent with 0.3$-$100 for X-ray detected Seyferts and LINERs in the same percentile range. One object, NGC~4625, has a long exposure ($\sim$55 ks) and a stringent X-ray upper limit of $10^{36.5}$~\ergs. We note that for this object, the spectroscopic classification from \citet{veron-cetty10} may be questionable. In \citet{moustakas10}, it was classified as a star-forming nucleus based on the \NII/\halpha\ ratio alone. 
%% non-detection in H II or Transition
The situation for \HII\ nuclei or transition objects is difficult to say.  As many or most of these objects---especially the \HII\ nuclei---reside in late-type or even bulgeless galaxies, they may very well lack a central black hole, and hence no X-ray core is expected in them.

%%%% -------- Table of Classes -------------------
%\floattable 
\begin{deluxetable}{lrcc}
  \tablewidth{0pc}
  \tablecaption{AGN numbers and detection rates as a function of optical spectral classification
   \label{tab:spectral_class}}
  \tablehead{ \colhead{Spectral Class}                       &   
\colhead{Galaxies}                      &   
\colhead{AGNs}                      &   
\colhead{Rate (\%)}                   }
  \startdata
  Seyfert 1 & 19 & 18 & 90$^{+ 7}_{- 12}$  \\
Seyfert 2 & 40 & 34 & 83$^{+ 8}_{- 10}$  \\
Seyfert 1 and 2 & 59 & 52 & 86$^{+ 6}_{- 7}$  \\
LINER 1 & 16 & 16 & 94$^{+ 5}_{- 10}$  \\
LINER 2 & 50 & 41 & 80$^{+ 8}_{- 9}$  \\
LINER 1 and 2 & 66 & 57 & 85$^{+ 6}_{- 7}$  \\
Transition & 41 & 28 & 67$^{+ 11}_{- 12}$  \\
H \textsc{ii} nucleus & 163 & 51 & 31$^{+ 6}_{- 5}$  \\
Absorption-line & 89 & 55 & 61$^{+ 8}_{- 8}$  \\
All & 418 & 243 & 58$^{+ 3}_{- 3}$  \\

  \enddata
\label{tab:agnclass}
\tablecomments{Errors are quoted at 90\% confidence level.}%with the Bayesian approach.}
\end{deluxetable}

%% ------ BAR/bar
\subsection{The Influence of Bars}
\label{subsec:bar}

\begin{figure}[t]
  \centering
  \includegraphics[width=0.7\columnwidth]{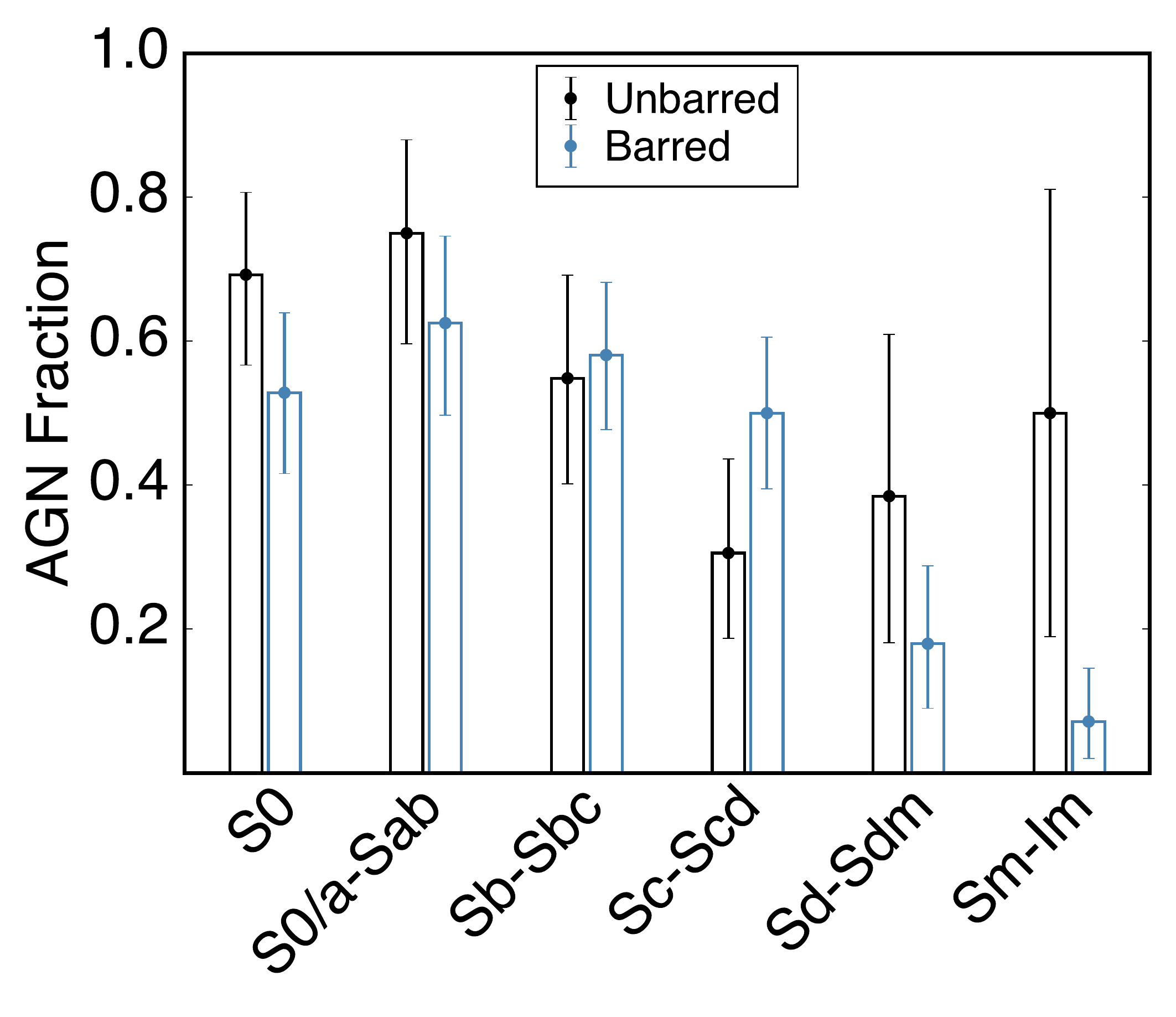}
  \caption{AGN detection rates for unbarred (SA; black) and barred (SAB and SB; blue) galaxies. Errors are quoted at 90\% confidence level.
  \label{fig:agnrate_bar}}
\end{figure}

Among disk galaxies, those without a bar have a marginally higher detection rate of AGN candidates compared to barred (SAB or SB) systems: $54^{+6}_{-6}$\% versus $43^{+4}_{-4}$\%. This is the exact opposite of what is expected, if bars, as often supposed (e.g., \citealt{heller94}; see a review in \citealt{kormendy04}), help deliver gas to the center of a galaxy.  However, when studying the possible correlations between nuclear activity and a large-scale bar, the dependence of bar frequency on Hubble type should be taken into account, as bars are more prevalent in late-type spirals. To account for this effect, we compare the X-ray AGN detection rates and their uncertainties (given at a 90\% confidence level) of barred and unbarred galaxies separated by Hubble type (Figure~\ref{fig:agnrate_bar}; Table~\ref{tab:agnrates}). For galaxies with Hubble types S0$-$Sab or Sd$-$Im, the X-ray AGN detection rates are slightly higher for unbarred galaxies than barred galaxies, whereas for Hubble types Sb$-$Scd, unbarred galaxies have marginally lower X-ray AGN rates than barred galaxies.  
We note that the error bars in Figure~\ref{fig:agnrate_bar} are relative large, and  any apparent differences in AGN fractions for barred and unbarred are not statistically significant.  We find no compelling evidence that bars influence AGN activity, at least not within the context of the current sample of mostly low-luminosity sources in nearby galaxies.  Similarly, \citet{ho97c} found that bars have a negligible effect on optically-selected AGN activity in the Palomar sample. 

%%%% -------- Figure of Hubble types and Bar -----------

\section{Black Hole Mass and Eddington Ratio}
\label{sec:mbh_edd}

\begin{figure*}[t]
  \centering
  \includegraphics[width=0.45\textwidth]{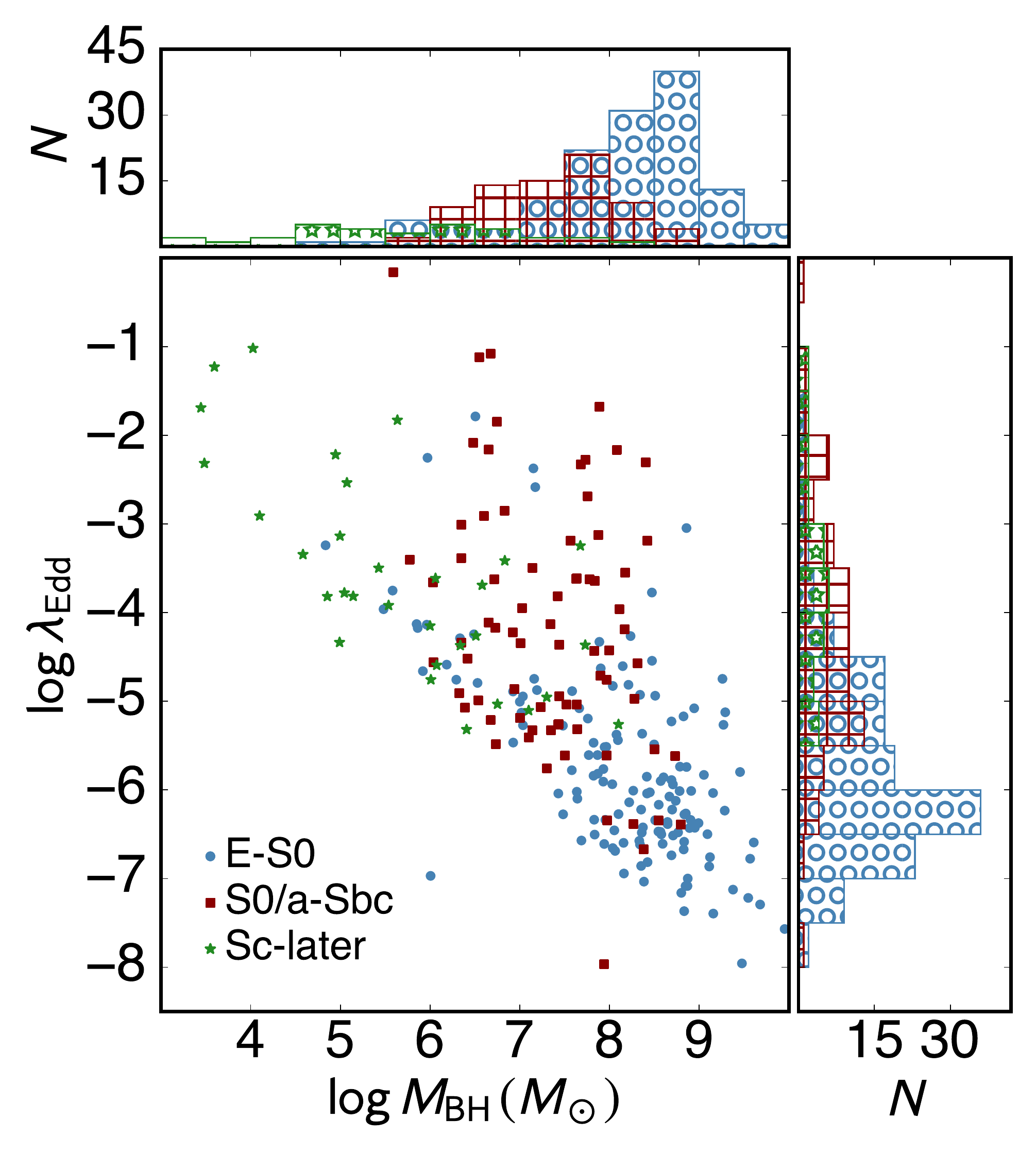}
  \includegraphics[width=0.45\textwidth]{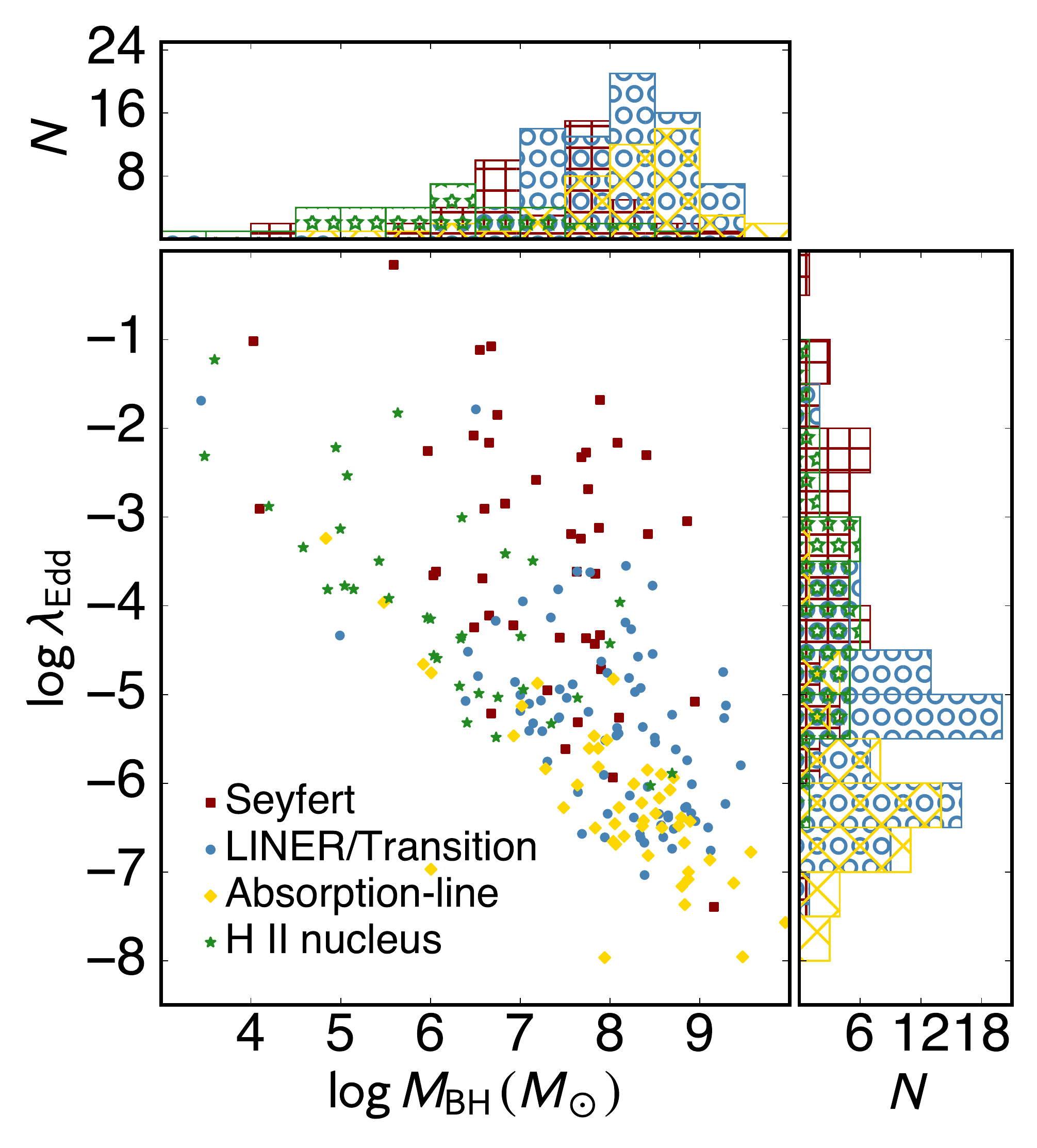}
  \caption{Distributions of \mbh\ and \eddr\ for galaxies binned by Hubble type and spectral classification.
  \label{fig:mbh}}
\end{figure*}

Paper I describes our methodology for deriving black hole masses (\mbh), X-ray luminosities (\lx; 2$-$10~keV), and Eddington ratios (\eddr\ $\equiv C_{\rm X}\cdot$ \lx/\ledd; $C_{\rm X} = 16$ adopted). The statistics of these properties  are summarized in Table~\ref{tab:stat} for the sample as a whole, as well as for various subgroups divided by Hubble type and optical spectral classification.  Figure~\ref{fig:mbh} plots the distributions of \mbh\ and \eddr.

%%%% -------- Table of Statistics of Distributions --------------
%\floattable 
\begin{deluxetable*}{lcccccccccccc}
\tabletypesize{\scriptsize}
\tablewidth{0pt}
\tablecaption{AGN Candidate Statistics}
\tablehead{
\colhead{Sample} & 
\multicolumn{4}{c}{$\log$ \mbh\ (\solarmass)} &
\multicolumn{4}{c}{$\log$ \lx\ (\ergs)} &
\multicolumn{4}{c}{$\log$ \eddr} \\ 
\cmidrule(lr){2-5}  \cmidrule(lr){6-9}  \cmidrule(l){10-13} 
                             &
\colhead{$N$}                &
\colhead{Median}                   &
\colhead{Min}                      &
\colhead{Max}                      &
\colhead{$N$}                &
\colhead{Median}                   &
\colhead{Min}                      &
\colhead{Max}                      &
\colhead{$N$}                      &
\colhead{Median}                   &
\colhead{Min}                      &
\colhead{Max}                      }
\startdata
%% Class/Type & Num & Median & Min & Max & ------ 
 All            &  250 & 7.83 & 3.45 & 9.95 &  314 & 39.26 & 35.94 & 43.11 &  250 & $-$5.18 & $-$7.97 & $-$0.16 \\ 
\\
 E              &   75 & 8.55 & 4.84 & 9.61 &   80 & 39.08 & 35.94 & 41.60 &   75 & $-$6.17 & $-$7.96 & $-$3.24 \\ 
 S0             &   64 & 8.07 & 5.58 & 9.95 &   70 & 39.27 & 36.56 & 42.71 &   64 & $-$5.65 & $-$7.57 & $-$1.79 \\ 
 S0/a-Sab       &   35 & 7.64 & 6.35 & 8.80 &   47 & 39.50 & 38.45 & 43.11 &   35 & $-$4.35 & $-$6.67 & $-$1.08 \\ 
 Sb-Sbc         &   40 & 7.12 & 5.59 & 8.55 &   53 & 39.98 & 36.88 & 42.36 &   40 & $-$4.35 & $-$7.97 & $-$0.16 \\ 
 Sc-later       &   31 & 5.64 & 3.45 & 8.10 &   54 & 38.92 & 36.79 & 42.04 &   31 & $-$3.78 & $-$5.32 & $-$1.02 \\ 
\\
 Seyfert        &   44 & 7.54 & 4.03 & 9.16 &   52 & 40.61 & 37.87 & 43.11 &   44 & $-$3.22 & $-$7.39 & $-$0.16 \\ 
 LINER          &   51 & 8.26 & 6.39 & 9.46 &   57 & 39.56 & 38.02 & 41.60 &   51 & $-$5.37 & $-$7.03 & $-$3.55 \\ 
 Transition     &   28 & 7.84 & 3.45 & 8.95 &   28 & 39.17 & 37.56 & 41.62 &   28 & $-$5.32 & $-$6.74 & $-$1.69 \\ 
 \HII           &   35 & 6.32 & 3.48 & 8.69 &   51 & 38.87 & 36.56 & 41.05 &   35 & $-$4.14 & $-$6.03 & $-$1.23 \\ 
 Absorption-line &   49 & 8.27 & 4.84 & 9.95 &   55 & 38.98 & 35.94 & 40.91 &   49 & $-$6.34 & $-$7.97 & $-$3.24 \\ 

\enddata
\label{tab:stat}
\tablecomments{Black hole masses are estimated from the \msigma\ relation (see details in Paper I). \lx\ is the X-ray luminosity in the 2$-$10 keV band. Eddington ratios (\eddr) are based on the X-ray luminosity and a bolometric correction factor of $C_{\rm X}$=16. }
\end{deluxetable*}

%%%% -------- Figures of Mbh-Edd scatter  --------------
%% ---- Lx, in types and classes

The 2$-$10 keV X-ray luminosities of the AGN candidates in this work lie in the range \lx\  $\approx 10^{36}-10^{43}$~\ergs, with a median value of $10^{39.3}$~\ergs\ (Table~\ref{tab:stat}), which indicates most of these AGN candidates are in the regime of low-luminosity AGNs. The most luminous of the nuclei are found in early-type spirals, whereas galaxies with Hubble types Sc and later have the least luminous nuclei. Seyferts and LINERs are more luminous than \HII\ nuclei, and \lx\ decreases along the sequence Seyferts $\rightarrow$ LINERs $\rightarrow$ transition objects $\rightarrow$ \HII\ nuclei.  This result, drawn entirely from our new X-ray measurements, mirrors the activity sequence proposed by \citet{ho09}, which was based on \halpha\ luminosities and a set of largely independent X-ray data. 

%% general correlations??
As previously noted \citep{ho08,ho09,gallo10,miller12}, the nuclei of nearby galaxies generally exhibit an inverse correlation between the Eddington ratio and  black hole mass  (Figure~\ref{fig:mbh}).   This is due, in part, to the fact that black holes with higher mass can be detected more easily.
%% Mbh, in types and classes
The black hole masses in this work cover a very large range, from $\sim3\times10^3$ to $9\times10^9$ \solarmass\, with a median value of $7\times10^7$ \solarmass. As expected, late-type (Sc and later) galaxies have the smallest black hole masses, while elliptical galaxies have the largest. As the majority of late-type galaxies have a spectral classification (if available) of \HII\ nuclei, it is not surprising that \HII\ nuclei have the smallest black hole masses, while LINERs and absorption-line nuclei occupy the upper end of the mass distribution.   We note that the uncertainties of the black hole mass estimates, derived from the  \msigma\ relation, can be considerable. \citet{kormendy13} show that  classical bulges and ellipticals obey a tight \msigma\ relation with an intrinsic scatter of only 0.28~dex.  By contrast, the \msigma\ relation of pseudobulges---those with a protracted evolutionary history driven by secular, internal processes---exhibits much larger scatter and a significantly lower zero point compared to classical bulges \citep[see also][]{greene16}.  A practical difficulty we face is that the vast majority of the galaxies in our sample lack reliable bulge classifications, and hence we cannot use a bulge-type--specific \msigma\ relation to estimate black hole masses.  Under these circumstances, we have little choice but to derive a single \msigma\ relation for all bulge types (pseudobulges combined with classical bulges and ellipticals), which has a larger intrinsic scatter of 0.44~dex (Paper I).  As the typical uncertainty of the bulge stellar velocity dispersions is only 5\%--15\% \citep{ho09b}, the final uncertainties of the black hole masses are dominated by the intrinsic scatter of the \msigma\ relation.  Moreover, because the zero point of the \msigma\ relation of pseudobulges is lower than that of classical bulges and ellipticals, using a single, combined \msigma\ relation will tend to systematically overestimate the black hole masses for pseudobulges and bulgeless galaxies, while ellipticals and classical bulges will have their black hole masses systematically underestimated.   Conversely, Eddington ratios will be underestimated in pseudobulges but overestimated in classical bulges and ellipticals.

%% Eddington ratios, in types and classes
The Eddington ratios spread over an even larger dynamic range, from $\sim10^{-8}$ to $\sim0.7$ , with a median value of $\sim7\times10^{-6}$.  Our sample contains  predominantly black holes accreting at extremely sub-Eddington rates. The AGN candidates in ellipticals, having the most massive black holes, exhibit the lowest values of \eddr, whereas late-type galaxies (Sc and later) have relatively higher \eddr\ as a result of their lower black hole masses. Among objects traditionally recognized to host black holes, at a given black hole mass, Seyferts, as a class, have the  highest Eddington ratios (median $10^{-3.2}$), followed by LINERs and transition objects (median $\sim 10^{-5.3}$), and then absorption-line nuclei (median $10^{-6.3}$).  The AGN candidates newly identified in \HII\ nuclei, with median \eddr$=10^{-4.1}$, lie in between Seyferts and LINERs/transition objects.

Apart from the new population of AGN candidates in \HII\ nuclei, the results above are broadly consistent with those previously obtained from optical observations \citep{ho97b,ho03}.  \citet{ho09}, using a compilation of more heterogeneous and generally less sensitive X-ray measurements from various literature sources, also derived very similar statistical results for the X-ray luminosities and Eddington ratios of the galaxies in the Palomar survey. This investigation, however, represents a significant improvement in sample size and data quality, as all of the X-ray data come from \Chandra/ACIS observations.  It is clear that---with sufficiently high sensitivity and angular resolution---X-ray observations
%%% ------  some discussion on X-ray more powerful ---------
allow us to detect nuclear accretion at much lower levels of activity than possible by optical techniques. In this work, for AGN candidates in \HII\ and absorption-line nuclei, the median value of X-ray luminosity is \lx $\simeq 10^{38.9}$~\ergs, which corresponds to \lha $\simeq 10^{37.8}$~\ergs,\footnote{Low-luminosity AGNs have a bolometric correction of \lbol $\simeq$ 220\lha $\simeq$ 16\lx\ \citep{ho08}} about an order of magnitude lower than the median \lha\ for the Palomar objects \citep[][$\sim10^{38.6}$~\ergs]{ho97b}.  In addition to sheer sensitivity, \Chandra/ACIS observations, which have the unique capability of ``filtering out" most of the contamination from host galaxy starlight, also afford more reliable detections of faint AGNs, which can be easily masked or confused by dust and star-forming regions at other wavelengths.  This is well illustrated in Figure \ref{fig:latetype} \citep[see also][]{ho01}.  The next section highlights a population of AGN candidates that was previously hidden among optically classified \HII\ nuclei.

\section{AGNs Candidates in \HII\ Nuclei} 
\label{sec:hii}

\HII\ nuclei, those whose dominant source of excitation derives from photoionization by young, massive stars, are extremely prevalent.  The Palomar optical spectroscopic survey of nearby galaxies found that 40\% of all nearby galaxies host an \HII\ nucleus, and for late-type galaxies the fraction is even higher, reaching 80\% \citep{ho97,ho97b}.  Compared to \HII\ regions in galactic disks, however, \HII\ nuclei exhibit systematically stronger low-ionization optical forbidden lines, suggesting that they contain an extra source of excitation not present in normal star-forming regions \citep{kennicutt89,ho97d}.  While \citet{kennicutt89} suggested that the additional excitation may come from a hidden AGN, \citet{ho97d} argued that shocks are more likely, although they noted that X-ray observations would be effective in uncovering optically hidden AGNs, should they be present.

\begin{figure*}[htbp]
  \centering
  \includegraphics[width=\textwidth]{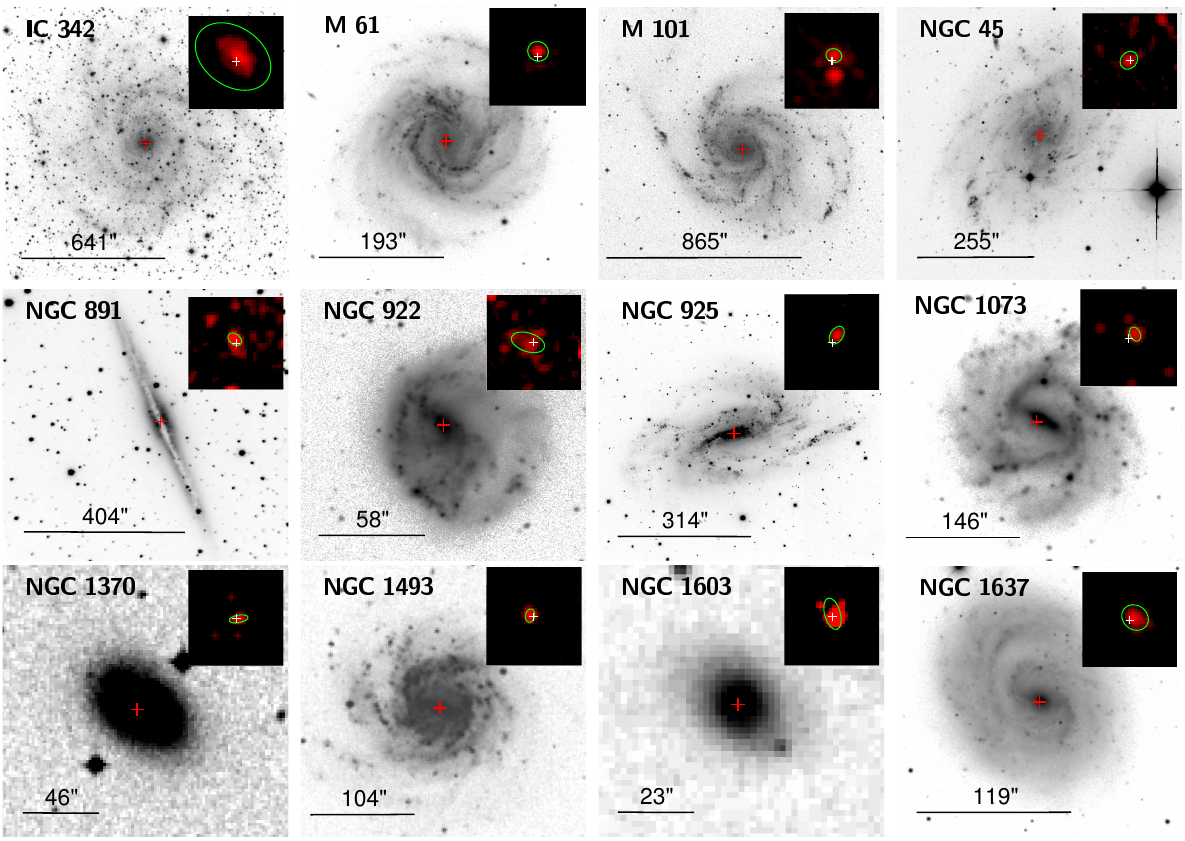}
  %smoothed with a gaussian of $\sigma=1$ pixel
  \caption{Optical and \Chandra/ACIS images (upper right corner) of AGN candidates in \HII\ nuclei. The horizontal segment in the bottom left corner of each panel represents a scale of $\frac{1}{2}D_{25}$ from RC3, and each X-ray image subtends $15\arcsec\times15\arcsec$, smoothed with a Gaussian of $\sigma=1$ pixel in {\tt ds9}. The cross marks the near-infrared/optical nuclear position, and the green ellipse in the X-ray image is the source region given by {\tt wavdetect}. North is up and east is to the left.
  \label{fig:latetype}}
\end{figure*}

\begin{figure*}[htbp]
  \centering
  \includegraphics[width=0.8\textwidth]{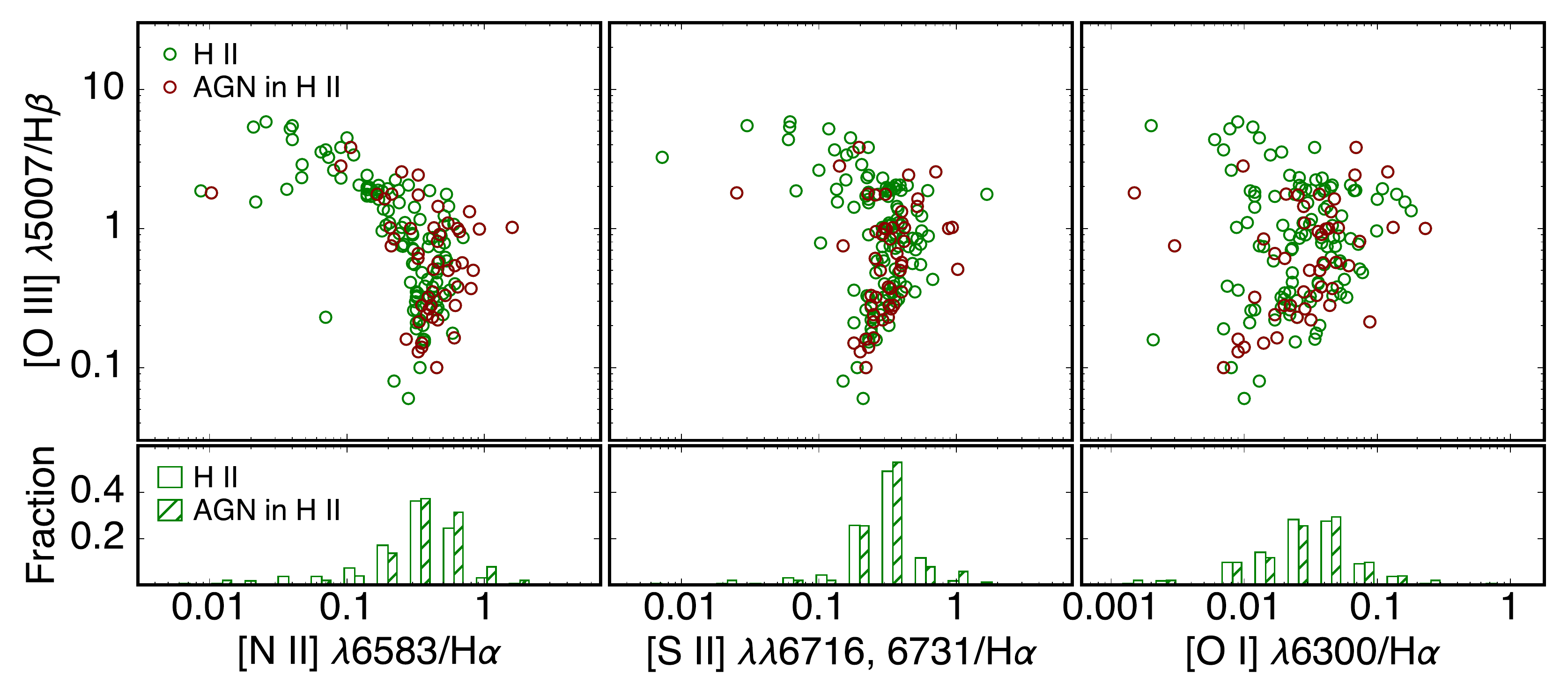}
  \caption{Optical emission-line flux ratio diagnostic diagrams for the sample of \HII\ nuclei in this study (upper panels) and histogram of strength of low-ionization lines (lower panels). 
  \label{fig:bpt}}
\end{figure*}

This work confirms that a significant fraction---31\% (51/\HiiNum)---of optically classified \HII\ nuclei have an X-ray core.
The general and X-ray properties of this special subsample of AGN candidates are listed in Table~\ref{tab:latetype}.  To highlight the morphology of the host galaxies, we assembled the highest available resolution optical ($B$ band if available, other bands otherwise) images from NED. These are displayed in Figure~\ref{fig:latetype}, superposed with the \Chandra/ACIS view of the central 15$\arcsec$$\times$15$\arcsec$ area to illustrate the clear association of the compact X-ray core with the stellar nucleus.  The very late-type morphologies of the host galaxies are immediately obvious, as is the fact that many of them clearly lack a bulge altogether.
Figure~\ref{fig:bpt} examines the \citet{veilleux87} optical line-intensity ratio diagnostic diagrams for the sample of \HII\ nuclei included in this study.  It is interesting that the subset with detected X-ray cores do {\it not}\ show preferentially enhanced low-ionization lines of \NII, \SII, or \OI.  
Thus, while some \HII\ nuclei do harbor optically hidden AGNs, they are sufficiently weak that they do not contribute significantly to the ionization budget of the optical line emission, and, in particular, they do not appear to be the primary cause for the excitation enhancement reported by \citet{kennicutt89} and \citet{ho97d}.

%% ----------------------------------------------
%%  Table of Late Type galaxies
\begin{deluxetable*}{llrccrrr}
\tablewidth{0pc} \renewcommand{\arraystretch}{1.2}
\tablecaption{Properties of AGN Candidates in \HII\ Nuclei \label{tab:latetype}}
\tablehead{\colhead{Name} & \colhead{Hubble Type} & \colhead{$\log$ \mbh} & \colhead{$\log$ \lha} & \colhead{Model} & \colhead{$\Gamma$} & \colhead{\nhint} & \colhead{$\log$ \lx}\\ \colhead{ } & \colhead{ } & \colhead{(\solarmass)} & \colhead{(\ergs)} & \colhead{ } & \colhead{ } & \colhead{($10^{22}$ cm$^{-2}$)} & \colhead{(\ergs)}}
\startdata
IC 342 & $\rm{SAB(rs)cd}$ & 6.07$_{-0.38}^{+0.32}$ & 38.11 & PL+VME & 1.92$_{-0.25}^{+0.21}$ & \nodata & 38.38$_{-0.05}^{+0.05}$ \\
M 61 & $\rm{SAB(rs)bc}$ & 6.35$_{-0.08}^{+0.08}$ & 39.93 & PL+ME & 1.53$_{-0.71}^{+0.46}$ & \nodata & 38.91$_{-0.32}^{+0.37}$ \\
M 101 & $\rm{SAB(rs)cd}$ & 3.48$_{-1.04}^{+0.71}$ & 39.14 & PL & 1.88$_{-0.20}^{+0.20}$ & \nodata & 38.07$_{-0.16}^{+0.14}$ \\
NGC 45 & $\rm{SA(s)dm}$ & \nodata & \nodata & \nodata & \nodata & $<$1.60 & 37.22$_{-0.67}^{+0.54}$ \\
NGC 891 & $\rm{SA(s)b?\ }$ & 6.04$_{-0.34}^{+0.29}$ & 38.85L & \nodata & \nodata & 9.04$_{-2.89}^{+17.31}$ & 38.38$_{-0.37}^{+0.67}$ \\
NGC 922 & $\rm{SB(s)cd\ pec}$ & \nodata & \nodata & \nodata & \nodata & $<$0.24 & 39.02$_{-0.40}^{+0.28}$ \\
NGC 925 & $\rm{SAB(s)d}$ & 6.00$_{-nan}^{+nan}$ & 38.36 & \nodata & \nodata & $<$0.45 & 38.75$_{-0.46}^{+0.35}$ \\
NGC 1073 & $\rm{SB(rs)c}$ & 3.60$_{-0.98}^{+0.68}$ & 38.36 & \nodata & \nodata & 0.57$_{-0.54}^{+0.83}$ & 39.27$_{-0.37}^{+0.40}$ \\
NGC 1370 & $\rm{cD?}$ & 5.96$_{-0.08}^{+0.08}$ & \nodata & \nodata & \nodata & $<$0.52 & 38.73$_{-0.52}^{+0.40}$ \\
NGC 1493 & $\rm{SB(r)cd}$ & \nodata & \nodata & \nodata & \nodata & $<$0.32 & 38.41$_{-0.36}^{+0.26}$ \\
NGC 1603 & $\rm{E?}$ & 7.03$_{-0.10}^{+0.10}$ & \nodata & \nodata & \nodata & $<$0.13 & 38.99$_{-0.43}^{+0.30}$ \\
NGC 1637 & $\rm{SAB(rs)c}$ & \nodata & \nodata & PL & 3.01 & \nodata & 38.27$_{-0.31}^{+0.23}$ \\
NGC 1808 & $\rm{(R)SAB(s)a}$ & 7.64$_{-0.15}^{+0.14}$ & \nodata & PL+VME & 0.74$_{-0.39}^{+0.31}$ & \nodata & 39.50$_{-0.10}^{+0.10}$ \\
NGC 2276 & $\rm{SAB(rs)c}$ & 6.34$_{-0.27}^{+0.24}$ & 40.83 & \nodata & \nodata & $<$0.02 & 38.87$_{-0.36}^{+0.26}$ \\
NGC 2500 & $\rm{SB(rs)d}$ & 5.04$_{-1.43}^{+0.87}$ & 38.00 & \nodata & \nodata & $<$0.83 & 38.17$_{-0.40}^{+0.61}$ \\
NGC 2748 & $\rm{SAbc}$ & 6.32$_{-0.23}^{+0.21}$ & 39.99 & \nodata & \nodata & $<$0.35 & 38.32$_{-0.43}^{+0.30}$ \\
NGC 2782 & $\rm{SAB(rs)a\ pec}$ & 8.11$_{-0.12}^{+0.11}$ & 41.51 & PABS*PL+ME+G & 0.55$_{-0.43}^{+0.42}$ & 54.27$_{-37.43}^{+122.70}$ & 41.05$_{-0.37}^{+0.66}$ \\
NGC 2798 & $\rm{SB(s)a\ pec}$ & 7.01$_{-0.60}^{+0.48}$ & \nodata & PL+ME & 2.62$_{-0.25}^{+0.25}$ & \nodata & 39.56$_{-0.15}^{+0.12}$ \\
NGC 2993 & $\rm{Sa\ pec}$ & \nodata & \nodata & PL+ME & 1.37$_{-0.28}^{+0.33}$ & $<$0.14 & 40.03$_{-0.12}^{+0.11}$ \\
NGC 3077 & $\rm{I0\ pec}$ & 4.20$_{-1.43}^{+0.87}$ & 38.93 & PL & 1.69$_{-0.57}^{+0.66}$ & 1.42$_{-0.68}^{+0.92}$ & 38.22$_{-0.09}^{+0.08}$ \\
NGC 3184 & $\rm{SAB(rs)cd}$ & 4.85$_{-0.52}^{+0.42}$ & 39.72 & \nodata & \nodata & $<$0.54 & 37.94$_{-0.26}^{+0.39}$ \\
NGC 3198 & $\rm{SB(rs)c}$ & 5.00$_{-0.47}^{+0.39}$ & 39.62 & \nodata & \nodata & $<$0.95 & 38.76$_{-0.27}^{+0.34}$ \\
NGC 3310 & $\rm{SAB(r)bc\ pec}$ & 6.35$_{-0.03}^{+0.03}$ & 40.53 & PL & 1.41$_{-0.15}^{+0.16}$ & 0.44$_{-0.09}^{+0.10}$ & 40.25$_{-0.04}^{+0.04}$ \\
NGC 3367 & $\rm{SB(rs)c}$ & 5.64$_{-0.41}^{+0.34}$ & 41.34 & PL+ME & 0.98$_{-0.15}^{+0.16}$ & \nodata & 40.71$_{-0.06}^{+0.06}$ \\
NGC 3521 & $\rm{SAB(rs)bc}$ & 7.35$_{-0.13}^{+0.12}$ & 39.10b & PL & 1.92$_{-0.48}^{+0.65}$ & $<$0.64 & 38.92$_{-0.15}^{+0.13}$ \\
NGC 3665 & $\rm{SA0^0(s)}$ & 8.69$_{-0.09}^{+0.09}$ & 39.75b & \nodata & \nodata & $<$0.75 & 39.70$_{-0.24}^{+0.37}$ \\
NGC 3877 & $\rm{SA(s)c?}$ & 6.41$_{-0.26}^{+0.23}$ & 40.29 & \nodata & \nodata & $<$0.11 & 37.99$_{-0.40}^{+0.29}$ \\
NGC 4038 & $\rm{SB(s)m\ pec}$ & \nodata & \nodata & \nodata & \nodata & $<$0.06 & 38.22$_{-0.38}^{+0.28}$ \\
NGC 4102 & $\rm{SAB(s)b?}$ & 8.00$_{-0.13}^{+0.12}$ & 41.50 & PL+ME+G & 1.52$_{-0.49}^{+0.50}$ & \nodata & 40.47$_{-0.24}^{+0.19}$ \\
NGC 4136 & $\rm{SAB(r)c}$ & 4.58$_{-0.58}^{+0.46}$ & 38.30 & \nodata & \nodata & 1.01$_{-0.81}^{+1.27}$ & 38.14$_{-0.42}^{+0.49}$ \\
NGC 4194 & $\rm{IBm\ pec}$ & 6.83$_{-0.62}^{+0.49}$ & \nodata & PL+VME & 1.44$_{-0.42}^{+0.42}$ & $<$0.66 & 40.32$_{-0.10}^{+0.09}$ \\
NGC 4217 & $\rm{Sb\ }$ & 6.54$_{-0.11}^{+0.11}$ & 38.84 & \nodata & \nodata & 2.59$_{-1.47}^{+2.39}$ & 38.45$_{-0.55}^{+0.52}$ \\
NGC 4490 & $\rm{SB(s)d\ pec}$ & 4.95$_{-0.50}^{+0.41}$ & 38.83 & PL & 2.40$_{-0.17}^{+0.18}$ & 0.92$_{-0.11}^{+0.13}$ & 39.63$_{-0.04}^{+0.04}$ \\
NGC 4526 & $\rm{SAB0^0?(s)}$ & 8.45$_{-0.09}^{+0.09}$ & 39.61 & PL & 1.10$_{-0.16}^{+0.31}$ & $<$0.13 & 39.33$_{-0.12}^{+0.10}$ \\
NGC 4559 & $\rm{SAB(rs)cd}$ & 5.14$_{-0.43}^{+0.36}$ & 38.58 & \nodata & \nodata & $<$0.59 & 38.23$_{-0.27}^{+0.47}$ \\
NGC 4561 & $\rm{SB(rs)dm}$ & \nodata & \nodata & PL & 1.45 & \nodata & 39.56$_{-0.18}^{+0.33}$ \\
NGC 4654 & $\rm{SAB(rs)cd}$ & 5.07$_{-0.47}^{+0.39}$ & 39.97 & \nodata & \nodata & $<$1.49 & 39.44$_{-0.44}^{+0.48}$ \\
NGC 4666 & $\rm{SABc?}$ & \nodata & \nodata & \nodata & \nodata & $<$1.09 & 39.04$_{-0.38}^{+0.49}$ \\
NGC 4670 & $\rm{SB0/a(s)\ pec?}$ & \nodata & \nodata & \nodata & \nodata & $<$0.76 & 39.28$_{-0.27}^{+0.38}$ \\
NGC 4900 & $\rm{SB(rs)c}$ & 5.54$_{-0.38}^{+0.32}$ & 39.48L & \nodata & \nodata & $<$0.62 & 38.52$_{-0.28}^{+0.55}$ \\
NGC 5102 & $\rm{SA0^-}$ & \nodata & \nodata & \nodata & \nodata & $<$0.68 & 36.56$_{-0.30}^{+0.49}$ \\
NGC 5248 & $\rm{SAB(rs)bc}$ & 6.73$_{-0.22}^{+0.20}$ & 39.76 & \nodata & \nodata & $<$0.77 & 38.15$_{-0.54}^{+0.53}$ \\
NGC 5253 & $\rm{pec}$ & \nodata & \nodata & \nodata & \nodata & $<$0.01 & 36.90$_{-0.33}^{+0.25}$ \\
NGC 5483 & $\rm{SA(s)c}$ & \nodata & \nodata & \nodata & \nodata & $<$1.14 & 39.03$_{-0.33}^{+0.42}$ \\
NGC 5775 & $\rm{SBc?\ }$ & 6.75$_{-0.17}^{+0.16}$ & 39.54 & \nodata & \nodata & $<$1.00 & 38.62$_{-0.32}^{+0.39}$ \\
NGC 6764 & $\rm{SB(s)bc}$ & \nodata & \nodata & PL+ME & 1.82$_{-0.36}^{+0.31}$ & \nodata & 40.05$_{-0.10}^{+0.08}$ \\
NGC 6946 & $\rm{SAB(rs)cd}$ & 5.43$_{-0.42}^{+0.35}$ & 40.14 & PL & 2.23$_{-0.27}^{+0.30}$ & 0.58$_{-0.17}^{+0.19}$ & 38.83$_{-0.07}^{+0.06}$ \\
NGC 7320 & $\rm{SA(s)d}$ & \nodata & \nodata & \nodata & \nodata & $<$0.86 & 38.40$_{-0.26}^{+0.35}$ \\
NGC 7714 & $\rm{SB(s)b?\ pec}$ & 7.14$_{-0.42}^{+0.36}$ & \nodata & PL & 1.62$_{-0.12}^{+0.12}$ & 0.16$_{-0.04}^{+0.04}$ & 40.55$_{-0.04}^{+0.04}$ \\
UGC 5720 & $\rm{Im\ pec?}$ & \nodata & \nodata & \nodata & \nodata & $<$0.01 & 39.14$_{-0.32}^{+0.24}$ \\
UGCA 166 & \nodata & \nodata & \nodata & PL & 1.72$_{-0.18}^{+0.19}$ & \nodata & 39.34$_{-0.12}^{+0.11}$
\enddata
\tablecomments{
        Column 1: Galaxy name. 
        Column 2: Hubble type from NED. 
        Column 3: Log of black hole masses estimated using the \msigma\ relation. 
        More details are in Paper I. 
        Column 4: Extinction corrected \halpha\ luminosity from the Palomar survey \citep{ho97,ho03}. 
        The typical uncertianties is 10-30\%, 
        and the letters ``b'',``c'',``L'' and ``u'' indicate an uncertainty of 30-50\%, 100\%,
        a 3$\sigma$ lower limit, and a 3$\sigma$ upper limit, respectively. 
        Column 5: Spectral fitting models, see Paper I for details. 
        Here ``PL'',``ME'',``VME'',``PABS'',``BB'',``G'' denote for a {\tt powerlaw}, 
        {\tt mekal}, {\tt vmekal}, {\tt pcfabs}, {\tt bbody}, 
        and {\tt gaussian} (for iron K$\alpha$ line) component in {\tt XSPEC}, respectively.
        Column 6: Photon index $\Gamma$. 
        Column 7: Intrinsic absorption from spectral fitting or hardness ratio otherwise.
        Column 8: X-ray luminosity in 2-10 keV band.
        } 
\end{deluxetable*}

%% ----------------------------------------------

%%% --------------- LARGE FIGURES --------------------------
%% ----- IMAGES OF LATE TYPE GALAXIES

\begin{figure*}[htbp]
  \centering
    \includegraphics[width=\textwidth]{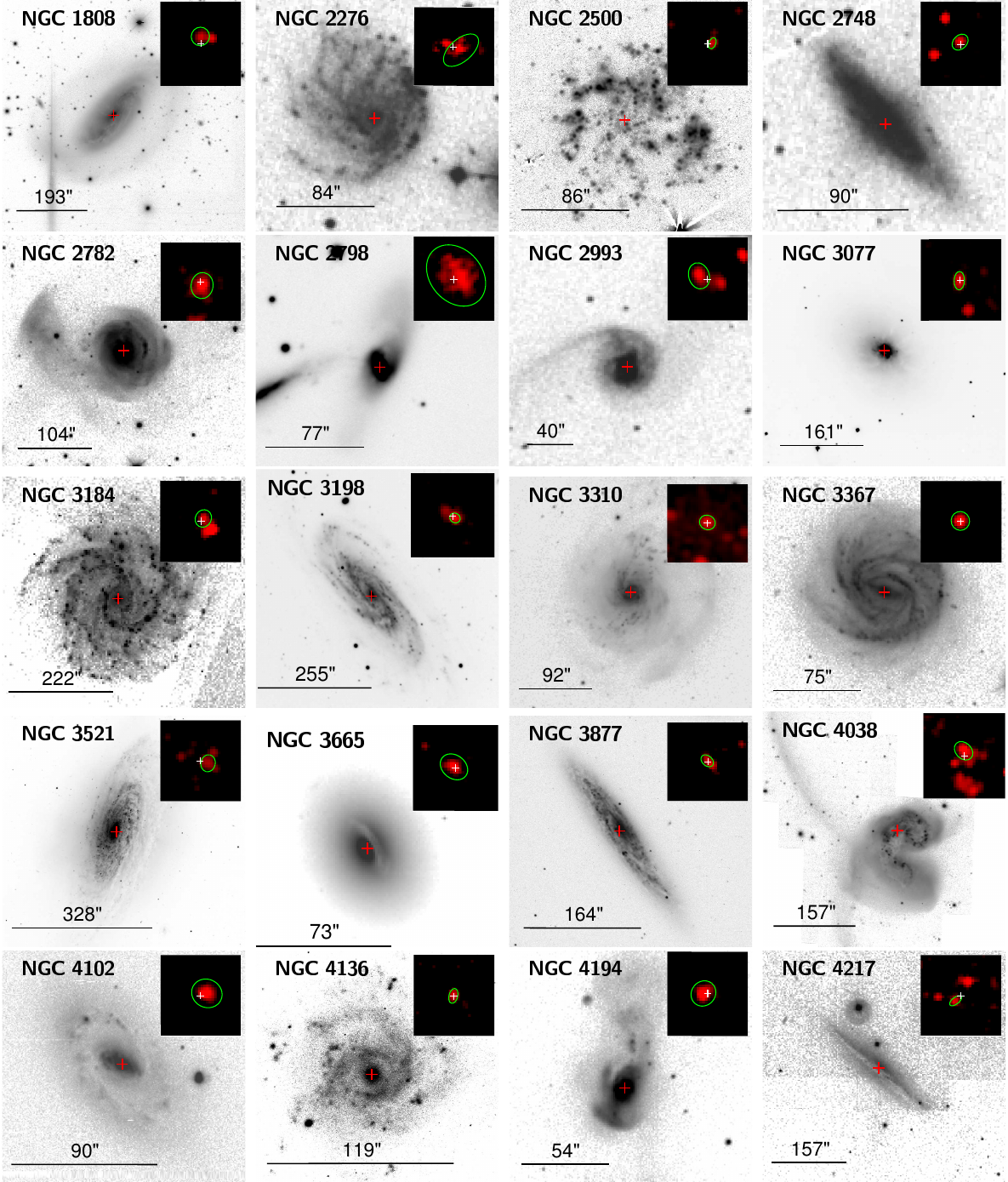}
  \centerline{Fig. \ref{fig:latetype} (Continued)}
\end{figure*}

\begin{figure*}[htbp]
  \centering
  \includegraphics[width=\textwidth]{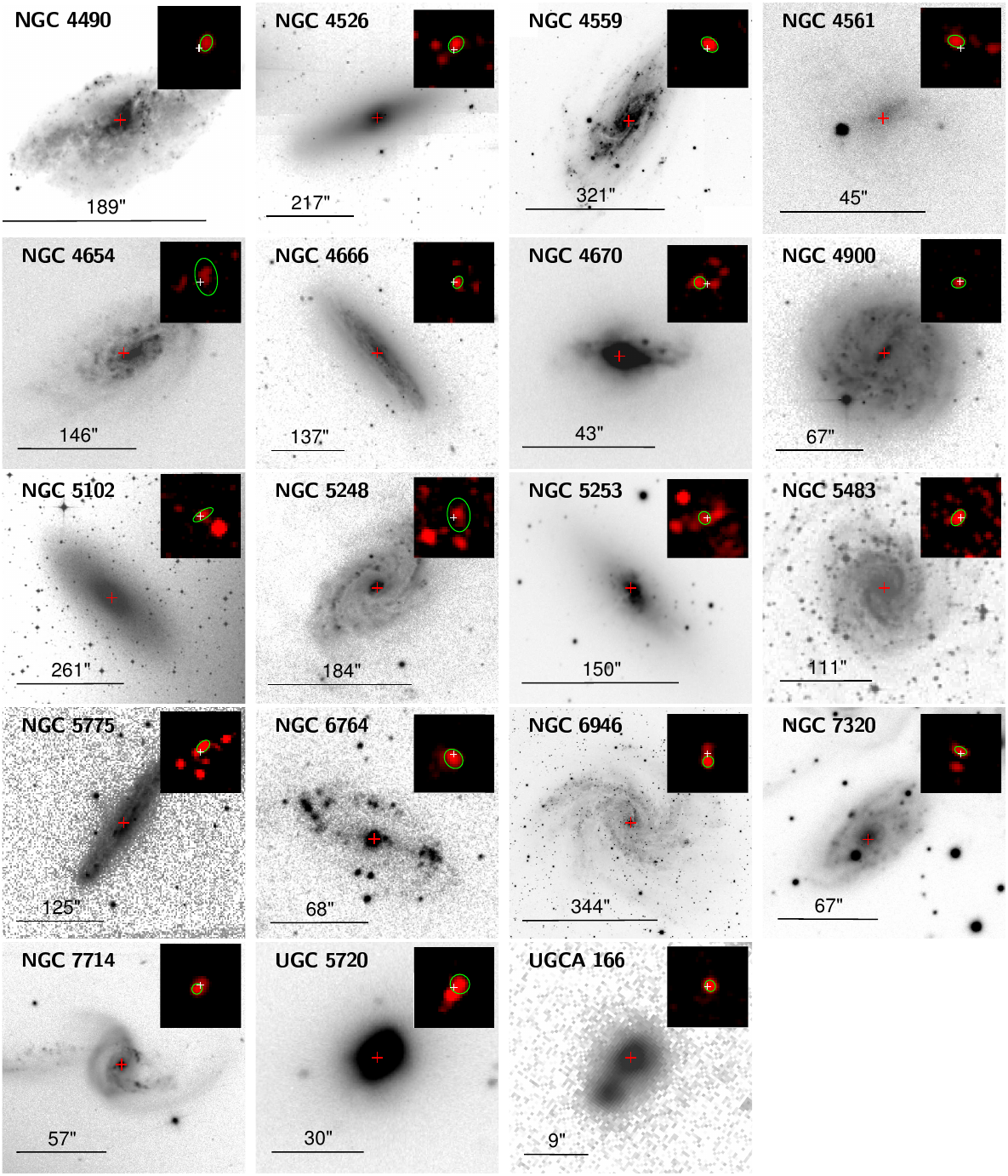}
  \centerline{Fig. \ref{fig:latetype} (Continued)}
\end{figure*}

%%% --------------- LARGE FIGURES --------------------------

%%% --------------- BPT Diagram --------------------------

Although we are cautious to call these ``candidate" AGNs, we have good reason to believe that this is the correct interpretation.
%% luminosity
With a median 2$-$10 keV luminosity of $\sim 8\times 10^{38}$~\ergs, 90\% of the AGN candidates in \HII\ nuclei exceeds $10^{38}$~\ergs. X-ray binaries associated with a potential nuclear star cluster may contaminate our AGN candidates. On the one hand, the luminosity function of low-mass X-ray binaries (LMXBs) shows a sharp cutoff at a few times $10^{38}$~\ergs\ and will have few sources above $10^{39}$~\ergs\ \citep{gilfanov04a}, suggesting that most of the sources above $10^{39}$~\ergs\ cannot be LMXBs. On the other hand, the typical age of $\sim$$10^{8-9}$~yr for nuclear star clusters \citep{rossa06} is too old to form high mass X-ray binaries (HMXBs), which could reach $\sim$$10^{40}$~\ergs\ \citep{swartz11a}, unless there is ongoing star formation. For the latter reason, \citet{desroches09a} argued that the HMXB contamination to the nuclear X-ray sources is negligible, and only the contribution from LMXBs is considered.  Here, to be conservative, we estimate both contributions to the detected X-ray cores. 

For HMXBs, their luminosity function scales with the star formation rate \citep{mineo12}, which itself depends on the \halpha\ luminosity, ${\rm SFR} \approx 10^{-41.27} \times$\ \lha~(\ergs)~\sfr \citep{kennicutt12}. Out of the 163 \HII\ nuclei, 90 of them have \lha\ measurements. For galaxies in \citet{mineo12}, \HII\ regions contribute $\sim$55\% of the total \lha\ emission \citep{greenawalt98,hoopes99a,thilker02}, and only $\sim$10\% of HMXBs (in 10$^{38}-10^{40}$~\ergs) are coincident with \HII\ regions (with position offsets less than $1\arcsec$).  As the \HII\ nuclei have \halpha\ luminosities not too different from that emitted by other \HII\ regions in the optical disk of these galaxies, we use \lha/0.55 to correct for the total \halpha\ luminosity in the whole galaxy, and renormalize the HMXB luminosity function by a factor of 0.1 to account for their association with the nuclear \HII\ region. 

\begin{figure}[t]
  \centering
  \includegraphics[width=0.8\columnwidth]{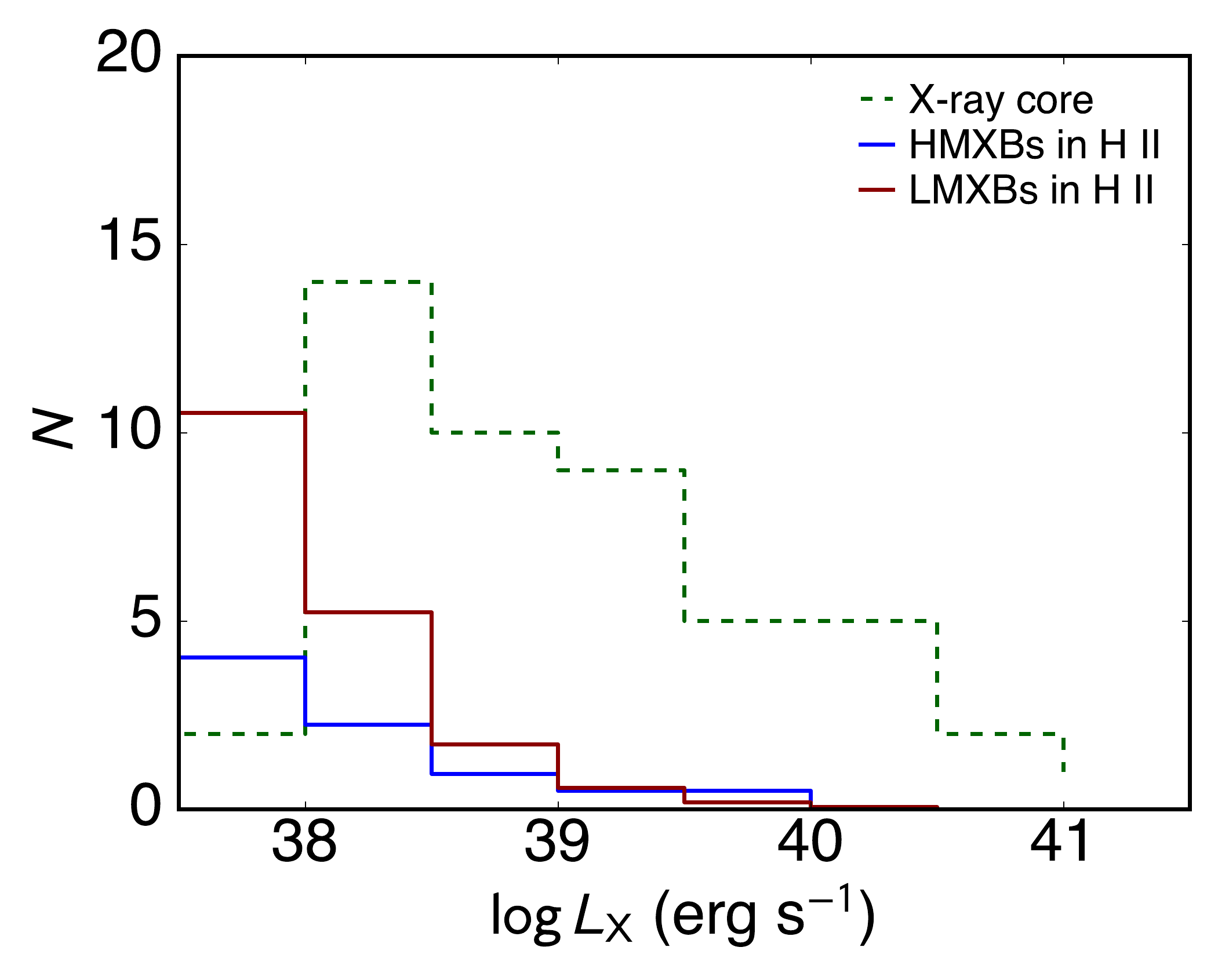}
  \caption{\lx\ distributions of AGN candidates in \HII\ nuclei and expected X-ray luminosity function of HMXBs and LMXBs.}
  \label{fig:xlf_hii}
\end{figure}

For LMXBs, their luminosity function is scaled with the total stellar mass. We treat nuclear clusters as globular clusters, as they share remarkably similar properties \citep{walcher05}. It is known that the LMXB production rate per unit stellar mass in globular clusters is roughly 100 times more efficient than in the field \citep{clark75,peacock17}. Moreover, metal-rich globular clusters are known to produce more LMXBs than metal-poor ones, roughly by a factor of 3 \citep{kundu07,kim13}. The metallicities of nuclear clusters in late-type galaxies are found to be similar to those of metal-rich clusters \citep{brodie06,rossa06}. Thus, we take the LMXB luminosity function in globular clusters from \citet{kim09} and the normalization from \citet{peacock17} as 0.11 LMXB with \lx$>2\times10^{37}$~\ergs\ per $10^{6}$~\solarmass\ for metal-rich globular clusters. The typical mass for the nuclear clusters in our sample ($\sim 2\times10^6$~\solarmass) is estimated from the 39 objects that are also shown in \citet{georgiev16}. The expected luminosity functions for both HMXBs and LMXBs are shown in Figure~\ref{fig:xlf_hii}.

Those constitute strong evidence that, in general, $\sim$40\% of the X-ray cores may be contaminated by X-ray binaries in the luminosity range of $10^{38}$--$10^{39}$~\ergs, and at luminosities above $10^{39}$~\ergs\ the contamination drops to 8\%.  The total contamination at luminosities of $10^{38}$~\ergs\ or higher is 26\%. Thus, the majority of the X-ray cores should be previously unrecognized low-luminosity AGNs, especially at luminosities above $10^{39}$~\ergs.

\begin{figure}[t]
  \centering
  \includegraphics[width=0.8\columnwidth]{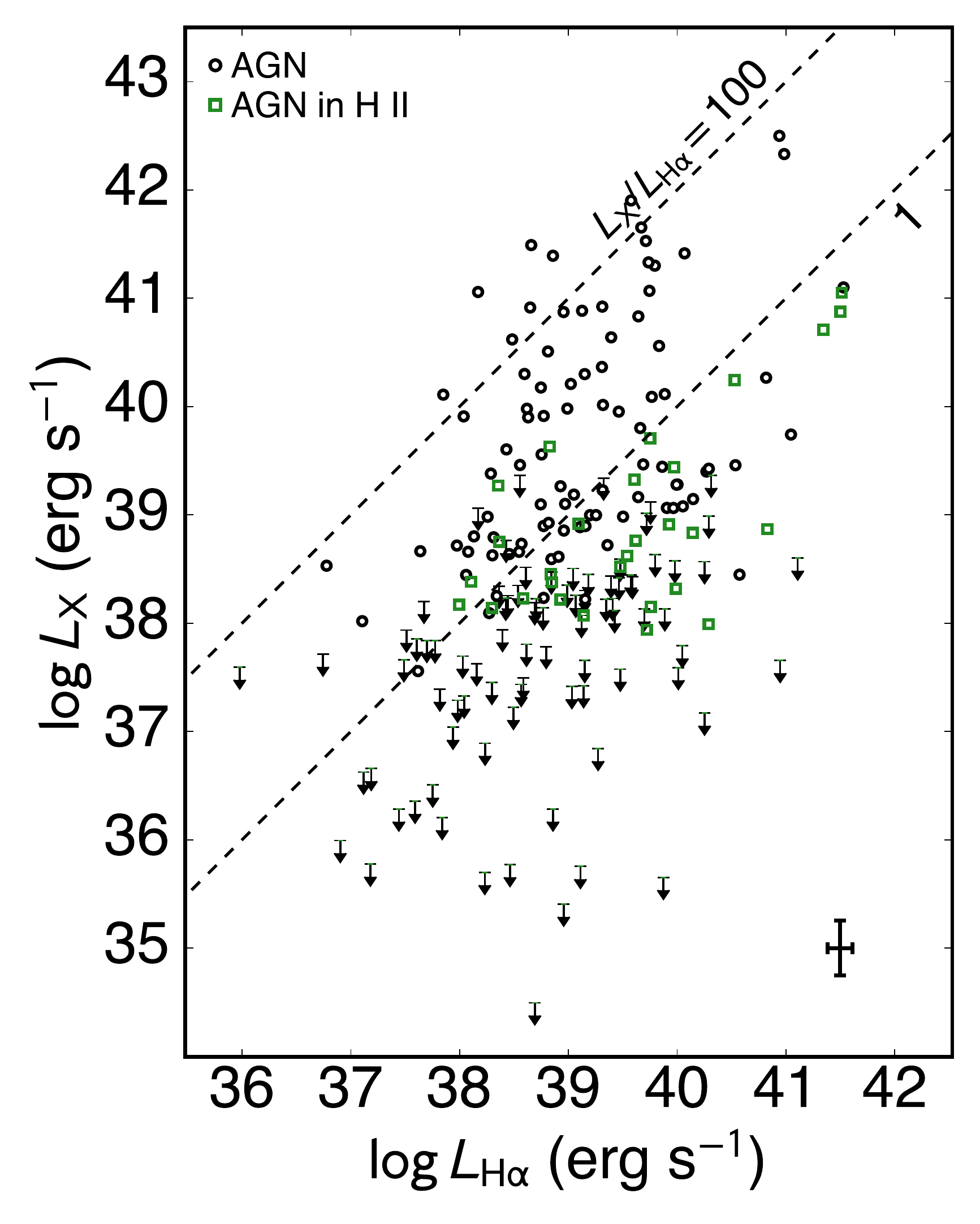}
  \caption{Correlation of X-ray luminosity and extinction-corrected H$\alpha$ luminosity. The black open circles denote X-ray AGN candidates, with upper limits marked by arrows.  Among the X-ray--detected galaxies,  those classified optically as \HII\ nuclei are marked with green squares. The dashed lines represent \lx/\lha=100 and 1.
The typical uncertainties of \lx\ and \lha\ are plot at the bottom right corner.
  \label{fig:halpha}}
\end{figure}

The ratios of \lx/\lha\ for the AGN candidates in \HII\ nuclei (with a median value of 0.3; see Figure~\ref{fig:halpha}) are considerably lower than those of other AGN candidates in Seyferts, LINERs or transition objects, which have a median \lx/\lha\ $\approx 5$.  A simple explanation for this order-of-magnitude difference in \lx/\lha\ is that the majority ($> 90$\%) of the \halpha\ emission from these \HII\ nuclei---as expected, given their optical classification---does not come from AGN photoionization but from star-forming activity.  The integrated optical emission from star-forming regions vastly overwhelm the weak AGN signal from these relatively low-mass central black holes.\footnote{Most of the optical classifications derive from observations from the Palomar \citep{ho95} or 6dF \citep{jones09} surveys, which employed an aperture size much larger than that of the X-ray observations.  The Palomar spectra were mostly extracted using a $2\arcsec \times 4\arcsec$ aperture, and the 6dF survey used a fiber of  diameter 6.7$\arcsec$; typical radius of the \chandra/ACIS point-spread function is $<1\arcsec$ for \chandra/ACIS.}

%% x-ray spectra
An important advantage of our study is that a sizable portion of our sample has enough counts to enable X-ray spectral analysis, and hence we can evaluate whether the X-ray sources have spectral properties consistent with other know low-luminosity AGNs \citep[e.g.,][]{ho08}.  Of the 21 (41\%) of AGN candidates associated with \HII\ nuclei for which we could analyze using {\tt XSPEC}, 11 can be well fitted with a single, absorbed power-law, nine requires a thermal component ({\tt mekal}), and one (NGC~2782) requires a partially covered absorption component ({\tt pcfabs}). Except for NGC~2782, 60\% of the photon indices  are in the range 1.4$-$2.0 with a typical value of of 1.7, which is only marginally flatter than the typical value for low-luminosity AGNs \citep{ho08}. Two sources (NGC~2782 and NGC~4102) show an iron K$\alpha$ emission line.

%% --- TIMING: for completeness ------
Lastly, timing analysis can, in principle, yield additional constraints on the physical origin of the X-ray emission. Unfortunately, due to the low count rates and short durations of most of the observations, the evidence based on timing is inconclusive.  We searched for variability on short timescales for the 11 sources in the sample with more than 500 photons in the 0.3$-$8 keV band. We analyzed their light curves with a time bin of 2000 s and calculated the $\chi^2/\rm dof$ for each light curve by assuming a constant count rate. Based on these light curves, the normalized excess variance \citep[$\sigma^2_{\rm NXS}$;][]{vaughan03} are also calculated. Only one object, NGC~3310, shows significant short timescale variability ($\chi^2/\rm dof$=76/24, $\sigma_{\rm NXS}=0.0305\pm 0.0133$); for other objects, only upper limits of $\sigma^2_{\rm NXS}$ can be given.  We also checked whether our \Chandra\ data can detect variability in some of the brighter,well-known AGNs in our full sample (e.g., NGC~4261, NGC~5506, NGC~4945, NGC~4051), whose short timescale variability has been previously studied in.   Some vary, but others do not. We note that the count rate and the total duration of the observations  are crucial for searching for short timescale variabilities.  The vast majority of the archival  \chandra\ data we used simply cannot yield meaningful constraints on X-ray variability.

\section{Implications for Low-mass Central Black Holes}

Most of the host galaxies of AGN candidates in \HII\ nuclei have late types (28 out of the 51 have Hubble types Sc$-$Im), whose bulge component, if it exists at all, is certainly not dominant. The AGN candidates in these \HII\ nuclei have the lowest median value of \mbh\ ($\sim 2\times10^6$~\solarmass) compared to other spectral classes (Seyferts, LINERs, transition objects and absorption-line nuclei; see Table~\ref{tab:stat}). As they occupy the low-mass end of the mass distribution of central black hole, these AGN candidates in \HII\ nuclei are very important for understanding the overall demography of ``seed'' supermassive black holes. Previous systematic searches for low-mass ($M_{\rm BH} < 10^6$~\solarmass) black holes based on optical selection \citep{greene04,greene07b,barth08,dong12,reines13,moran14} were only sensitive to sources radiating at a significant fraction of the Eddington limit.  It is highly uncertain whether these optically selected samples give an accurate census of  the incidence of low-mass black holes in the nearby universe.  This study, following the initial suggestion by \citet{desroches09a} based on a \Chandra\ study of a much more limited sample of late-type galaxies, confirms that the true frequency of low-mass black holes, dominated by objects accreting at sub-Eddington rates, is significantly higher.  Among the 51 detected X-ray \HII\ nuclei, 46 of them have an X-ray luminosity above $10^{38}$~\ergs. Taking into account the contamination from X-ray binaries ($\sim$26\% above $10^{38}$~\ergs), we regard $\sim$21\% as a reasonable {\it lower limit}\ on the true incidence of AGNs in late-type galaxies, as the contamination by X-ray binaries (both LMXBs and HMXBs) very likely has been overestimated, and some low-luminosity AGNs are not detected due to our sensitivity limit.

\section{Conclusions}
\label{sec:conclusion}
We identified \AGNNum\ AGN candidates out of \GalNum\ galaxies within 50 Mpc (Paper I). In this work, we study the demographics of AGNs in nearby galaxies, and reveal $\sim$50 hidden AGN candidates in \HII\ nuclei. The main conclusions are summarized below.

\begin{enumerate}[labelindent=0pt]
  \item In early-type galaxies, about 60\%--70\% of them are found to have an AGN candidate, and this fraction drops to less than 20\% for very late-type galaxies (Sdm and Sm). About 85\% of Seyferts and LINERs show an X-ray core. This fraction is $\sim$70\% for transition objects and $\sim$60\% for absorption-line nuclei. 30\% of \HII\ nuclei  have an X-ray core. We do not find a significant correlation between the presence of a large-scale bar and AGN activity.
  
  \item The X-ray luminosity decreases in a sequence of Seyferts $\rightarrow$ LINERs $\rightarrow$ transition objects $\rightarrow$ \HII\ nuclei. Late-type spirals have the lowest luminosities. 
  
  \item  In general, more massive black holes tend to display a smaller Eddington ratio. At a given black hole mass, the Seyferts have the highest Eddington level.
  
  \item In \HiiNum\ \HII\ nuclei, 51 out of them are identified to have an X-ray core. Most of these X-ray sources are not X-ray binaries. The morphology of these galaxies is of late-type and most likely has no or very small bulges. These objects are candidate low-mass black holes according to the \msigma\ relation, suggesting that the black hole occupation fraction in late-type galaxies is at least $\sim$21\%.

\end{enumerate}

\acknowledgments
The research of LCH is supported by the National Key Program for Science and Technology Research and Development (2016YFA0400702) and the National Science Foundation of China (11473002, 11303008).
HF acknowledges funding support from the National Natural Science Foundation of China under grant No.\ 11633003, and  the National Program on Key Research and Development Project (Grant No. 2016YFA040080X).

%\bibliographystyle{aasjournal}
%\bibliography{xraycores-new}

\begin{thebibliography}{}
\expandafter\ifx\csname natexlab\endcsname\relax\def\natexlab#1{#1}\fi

\bibitem[{Barth {et~al.}(2008)Barth, Greene, \& Ho}]{barth08}
Barth, A.~J., Greene, J.~E., \& Ho, L.~C. 2008, \aj, 136, 1179

\bibitem[{Brodie \& Strader(2006)}]{brodie06}
Brodie, J.~P., \& Strader, J. 2006, \araa, 44, 193

\bibitem[{Chen {et~al.}(2017)Chen, Brandt, Reines, Lansbury, Stern, Alexander,
  Bauer, Del~Moro, Gandhi, Harrison, Hickox, Koss, Lanz, Luo, Mullaney, Ricci,
  \& Trump}]{chen17}
Chen, C.-T.~J., Brandt, W.~N., Reines, A.~E., {et~al.} 2017, \apj, 837, 48

\bibitem[{Clark(1975)}]{clark75}
Clark, G.~W. 1975, \apjlett, 199, L143

\bibitem[{{de}~{Vaucouleurs} {et~al.}(1991){de}~{Vaucouleurs}, {de
  Vaucouleurs}, Corwin, Buta, Paturel, \& Fouqu{\'e}}]{devaucouleurs91}
{de}~{Vaucouleurs}, G., {de Vaucouleurs}, A., Corwin, H.~G., {et~al.} 1991,
  Third {{Reference Catalogue}} of {{Bright Galaxies}}. (Springer)

\bibitem[{Desroches \& Ho(2009)}]{desroches09a}
Desroches, L.-B., \& Ho, L.~C. 2009, \apj, 690, 267

\bibitem[{Dong {et~al.}(2012)Dong, Ho, Yuan, Wang, Fan, Zhou, \&
  Jiang}]{dong12}
Dong, X.-B., Ho, L.~C., Yuan, W., {et~al.} 2012, \apj, 755, 167

\bibitem[{Gallo {et~al.}(2008)Gallo, Treu, Jacob, Woo, Marshall, \&
  Antonucci}]{gallo08}
Gallo, E., Treu, T., Jacob, J., {et~al.} 2008, \apj, 680, 154

\bibitem[{Gallo {et~al.}(2010)Gallo, Treu, Marshall, Woo, Leipski, \&
  Antonucci}]{gallo10}
Gallo, E., Treu, T., Marshall, P.~J., {et~al.} 2010, \apj, 714, 25

\bibitem[{Georgiev {et~al.}(2016)Georgiev, B{\"o}ker, Leigh, L{\"u}tzgendorf,
  \& Neumayer}]{georgiev16}
Georgiev, I.~Y., B{\"o}ker, T., Leigh, N., L{\"u}tzgendorf, N., \& Neumayer, N.
  2016, \mnras, 457, 2122

\bibitem[{Gilfanov(2004)}]{gilfanov04a}
Gilfanov, M. 2004, \mnras, 349, 146

\bibitem[{Greenawalt {et~al.}(1998)Greenawalt, Walterbos, Thilker, \&
  Hoopes}]{greenawalt98}
Greenawalt, B., Walterbos, R. A.~M., Thilker, D., \& Hoopes, C.~G. 1998, \apj,
  506, 135

\bibitem[{Greene(2012)}]{greene12}
Greene, J.~E. 2012, {NatCo}, 3, 1304

\bibitem[{Greene \& Ho(2004)}]{greene04}
Greene, J.~E., \& Ho, L.~C. 2004, \apj, 610, 722

\bibitem[{Greene \& Ho(2007)}]{greene07b}
Greene, J.~E., \& Ho, L.~C. 2007, \apj, 670, 92

\bibitem[{Greene \& Ho(2009)}]{greene09}
Greene, J.~E., \& Ho, L.~C. 2009, \pasp, 121, 1167

\bibitem[{Greene {et~al.}(2016)Greene, Seth, Kim, L{\"a}sker, Goulding, Gao,
  Braatz, Henkel, Condon, Lo, \& Zhao}]{greene16}
Greene, J.~E., Seth, A., Kim, M., {et~al.} 2016, \apjlett, 826, L32

\bibitem[{Grier {et~al.}(2011)Grier, Mathur, Ghosh, \& Ferrarese}]{grier11}
Grier, C.~J., Mathur, S., Ghosh, H., \& Ferrarese, L. 2011, \apj, 731, 60

\bibitem[{Haehnelt \& Rees(1993)}]{haehnelt93}
Haehnelt, M.~G., \& Rees, M.~J. 1993, \mnras, 263, 168

\bibitem[{Heckman(1980)}]{heckman80a}
Heckman, T.~M. 1980, A\&A, 87, 152

\bibitem[{Heger {et~al.}(2003)Heger, Fryer, Woosley, Langer, \&
  Hartmann}]{heger03}
Heger, A., Fryer, C.~L., Woosley, S.~E., Langer, N., \& Hartmann, D.~H. 2003,
  \apj, 591, 288

\bibitem[{Heller \& Shlosman(1994)}]{heller94}
Heller, C.~H., \& Shlosman, I. 1994, \apj, 424, 84

\bibitem[{Ho(2008)}]{ho08}
Ho, L.~C. 2008, \araa, 46, 475

\bibitem[{Ho(2009)}]{ho09}
Ho, L.~C. 2009, \apj, 699, 626

\bibitem[{Ho {et~al.}(1993)Ho, Filippenko, \& Sargent}]{ho93}
Ho, L.~C., Filippenko, A.~V., \& Sargent, W. L.~W. 1993, \apj, 417, 63

\bibitem[{Ho {et~al.}(1995)Ho, Filippenko, \& Sargent}]{ho95}
Ho, L.~C., Filippenko, A.~V., \& Sargent, W. L.~W. 1995, \apjsupp, 98, 477

\bibitem[{Ho {et~al.}(1997{\natexlab{a}})Ho, Filippenko, \& Sargent}]{ho97c}
Ho, L.~C., Filippenko, A.~V., \& Sargent, W. L.~W. 1997{\natexlab{a}}, \apj,
  487, 591

\bibitem[{Ho {et~al.}(1997{\natexlab{b}})Ho, Filippenko, \& Sargent}]{ho97d}
Ho, L.~C., Filippenko, A.~V., \& Sargent, W. L.~W. 1997{\natexlab{b}}, \apj,
  487, 579

\bibitem[{Ho {et~al.}(1997{\natexlab{c}})Ho, Filippenko, \& Sargent}]{ho97}
Ho, L.~C., Filippenko, A.~V., \& Sargent, W. L.~W. 1997{\natexlab{c}},
  \apjsupp, 112, 315

\bibitem[{Ho {et~al.}(1997{\natexlab{d}})Ho, Filippenko, \& Sargent}]{ho97b}
Ho, L.~C., Filippenko, A.~V., \& Sargent, W. L.~W. 1997{\natexlab{d}}, \apj,
  487, 568

\bibitem[{Ho {et~al.}(2003)Ho, Filippenko, \& Sargent}]{ho03}
Ho, L.~C., Filippenko, A.~V., \& Sargent, W. L.~W. 2003, \apj, 583, 159

\bibitem[{Ho {et~al.}(2009)Ho, Greene, Filippenko, \& Sargent}]{ho09b}
Ho, L.~C., Greene, J.~E., Filippenko, A.~V., \& Sargent, W. L.~W. 2009,
  \apjsupp, 183, 1

\bibitem[{Ho {et~al.}(2001)Ho, Feigelson, Townsley, Sambruna, Garmire, Brandt,
  Filippenko, Griffiths, Ptak, \& Sargent}]{ho01}
Ho, L.~C., Feigelson, E.~D., Townsley, L.~K., {et~al.} 2001, \apjlett, 549, L51

\bibitem[{Hoopes {et~al.}(1999)Hoopes, Walterbos, \& Rand}]{hoopes99a}
Hoopes, C.~G., Walterbos, R. A.~M., \& Rand, R.~J. 1999, \apj, 522, 669

\bibitem[{Jones {et~al.}(2009)Jones, Read, Saunders, Colless, Jarrett, Parker,
  Fairall, Mauch, Sadler, Watson, Burton, Campbell, Cass, Croom, Dawe, Fiegert,
  Frankcombe, Hartley, Huchra, James, Kirby, Lahav, Lucey, Mamon, Moore,
  Peterson, Prior, Proust, Russell, Safouris, Wakamatsu, Westra, \&
  Williams}]{jones09}
Jones, D.~H., Read, M.~A., Saunders, W., {et~al.} 2009, \mnras, 399, 683

\bibitem[{Kennicutt \& Evans(2012)}]{kennicutt12}
Kennicutt, R.~C., \& Evans, N.~J. 2012, \araa, 50, 531

\bibitem[{Kennicutt {et~al.}(1989)Kennicutt, Keel, \& Blaha}]{kennicutt89}
Kennicutt, Jr., R.~C., Keel, W.~C., \& Blaha, C.~A. 1989, \aj, 97, 1022

\bibitem[{Kim {et~al.}(2013)Kim, Fabbiano, Ivanova, Fragos, Jord{\'a}n,
  Sivakoff, \& Voss}]{kim13}
Kim, D.-W., Fabbiano, G., Ivanova, N., {et~al.} 2013, \apj, 764, 98

\bibitem[{Kim {et~al.}(2009)Kim, Fabbiano, Brassington, Fragos, Kalogera,
  Zezas, Jord{\'a}n, Sivakoff, Kundu, Zepf, Angelini, Davies, Gallagher, Juett,
  King, Pellegrini, Sarazin, \& Trinchieri}]{kim09}
Kim, D.-W., Fabbiano, G., Brassington, N.~J., {et~al.} 2009, \apj, 703, 829

\bibitem[{Kormendy \& Ho(2013)}]{kormendy13}
Kormendy, J., \& Ho, L.~C. 2013, \araa, 51, 511

\bibitem[{Kormendy \& Kennicutt(2004)}]{kormendy04}
Kormendy, J., \& Kennicutt, Jr., R.~C. 2004, \araa, 42, 603

\bibitem[{Kundu {et~al.}(2007)Kundu, Maccarone, \& Zepf}]{kundu07}
Kundu, A., Maccarone, T.~J., \& Zepf, S.~E. 2007, \apj, 662, 525

\bibitem[{Lemons {et~al.}(2015)Lemons, Reines, Plotkin, Gallo, \&
  Greene}]{lemons15}
Lemons, S.~M., Reines, A.~E., Plotkin, R.~M., Gallo, E., \& Greene, J.~E. 2015,
  \apj, 805, 12

\bibitem[{Mezcua {et~al.}(2016)Mezcua, Civano, Fabbiano, Miyaji, \&
  Marchesi}]{mezcua16}
Mezcua, M., Civano, F., Fabbiano, G., Miyaji, T., \& Marchesi, S. 2016, \apj,
  817, 20

\bibitem[{Miller {et~al.}(2012)Miller, Gallo, Treu, \& Woo}]{miller12}
Miller, B., Gallo, E., Treu, T., \& Woo, J.-H. 2012, \apj, 747, 57

\bibitem[{Mineo {et~al.}(2012)Mineo, Gilfanov, \& Sunyaev}]{mineo12}
Mineo, S., Gilfanov, M., \& Sunyaev, R. 2012, \mnras, 419, 2095

\bibitem[{Moran {et~al.}(2014)Moran, Shahinyan, Sugarman, V{\'e}lez, \&
  Eracleous}]{moran14}
Moran, E.~C., Shahinyan, K., Sugarman, H.~R., V{\'e}lez, D.~O., \& Eracleous,
  M. 2014, \aj, 148, 136

\bibitem[{Mortlock {et~al.}(2011)Mortlock, Warren, Venemans, Patel, Hewett,
  McMahon, Simpson, Theuns, Gonz{\'a}les-Solares, Adamson, Dye, Hambly, Hirst,
  Irwin, Kuiper, Lawrence, \& R{\"o}ttgering}]{mortlock11}
Mortlock, D.~J., Warren, S.~J., Venemans, B.~P., {et~al.} 2011, Nature, 474,
  616

\bibitem[{Moustakas {et~al.}(2010)Moustakas, Kennicutt, Tremonti, Dale, Smith,
  \& Calzetti}]{moustakas10}
Moustakas, J., Kennicutt, Jr., R.~C., Tremonti, C.~A., {et~al.} 2010, \apjsupp,
  190, 233

\bibitem[{Nucita {et~al.}(2017)Nucita, Manni, De~Paolis, Giordano, \&
  Ingrosso}]{nucita17}
Nucita, A.~A., Manni, L., De~Paolis, F., Giordano, M., \& Ingrosso, G. 2017,
  \apj, 837, 66

\bibitem[{Pardo {et~al.}(2016)Pardo, Goulding, Greene, Somerville, Gallo,
  Hickox, Miller, Reines, \& Silverman}]{pardo16}
Pardo, K., Goulding, A.~D., Greene, J.~E., {et~al.} 2016, \apj, 831, 203

\bibitem[{Peacock {et~al.}(2017)Peacock, Zepf, Kundu, Maccarone, Lehmer,
  Gonzalez, \& Maraston}]{peacock17}
Peacock, M.~B., Zepf, S.~E., Kundu, A., {et~al.} 2017, \mnras, 466, 4021

\bibitem[{Reines \& Comastri(2016)}]{reines16a}
Reines, A.~E., \& Comastri, A. 2016, PASA, 33, e054

\bibitem[{Reines {et~al.}(2013)Reines, Greene, \& Geha}]{reines13}
Reines, A.~E., Greene, J.~E., \& Geha, M. 2013, \apj, 775, 116

\bibitem[{Rossa {et~al.}(2006)Rossa, {van der Marel}, B{\"o}ker, Gerssen, Ho,
  Rix, Shields, \& Walcher}]{rossa06}
Rossa, J., {van der Marel}, R.~P., B{\"o}ker, T., {et~al.} 2006, \aj, 132, 1074

\bibitem[{She {et~al.}(2017)She, Ho, \& Feng}]{she17}
She, R., Ho, L.~C., \& Feng, H. 2017, \apj, 835, 223

\bibitem[{Swartz {et~al.}(2011)Swartz, Soria, Tennant, \& Yukita}]{swartz11a}
Swartz, D.~A., Soria, R., Tennant, A.~F., \& Yukita, M. 2011, \apj, 741, 49

\bibitem[{Thilker {et~al.}(2002)Thilker, Walterbos, Braun, \&
  Hoopes}]{thilker02}
Thilker, D.~A., Walterbos, R. A.~M., Braun, R., \& Hoopes, C.~G. 2002, \aj,
  124, 3118

\bibitem[{Vaughan {et~al.}(2003)Vaughan, Edelson, Warwick, \&
  Uttley}]{vaughan03}
Vaughan, S., Edelson, R., Warwick, R.~S., \& Uttley, P. 2003, \mnras, 345, 1271

\bibitem[{Veilleux \& Osterbrock(1987)}]{veilleux87}
Veilleux, S., \& Osterbrock, D.~E. 1987, \apjsupp, 63, 295

\bibitem[{V{\'e}ron-Cetty \& V{\'e}ron(2010)}]{veron-cetty10}
V{\'e}ron-Cetty, M.-P., \& V{\'e}ron, P. 2010, A\&A, 518, A10

\bibitem[{Volonteri(2010)}]{volonteri10}
Volonteri, M. 2010, A\&A Review, 18, 279

\bibitem[{Volonteri {et~al.}(2008)Volonteri, Lodato, \&
  Natarajan}]{volonteri08}
Volonteri, M., Lodato, G., \& Natarajan, P. 2008, \mnras, 383, 1079

\bibitem[{Walcher {et~al.}(2005)Walcher, {van der Marel}, McLaughlin, Rix,
  B{\"o}ker, H{\"a}ring, Ho, Sarzi, \& Shields}]{walcher05}
Walcher, C.~J., {van der Marel}, R.~P., McLaughlin, D., {et~al.} 2005, \apj,
  618, 237

\bibitem[{Weisskopf {et~al.}(2002)Weisskopf, Brinkman, Canizares, Garmire,
  Murray, \& Van~Speybroeck}]{weisskopf02}
Weisskopf, M.~C., Brinkman, B., Canizares, C., {et~al.} 2002, \pasp, 114, 1

\bibitem[{Wu {et~al.}(2015)Wu, Wang, Fan, Yi, Zuo, Bian, Jiang, McGreer, Wang,
  Yang, Yang, Thompson, \& Beletsky}]{wu15}
Wu, X.-B., Wang, F., Fan, X., {et~al.} 2015, Nature, 518, 512

\bibitem[{Zhang {et~al.}(2009)Zhang, Soria, Zhang, Swartz, \& Liu}]{zhang09}
Zhang, W.~M., Soria, R., Zhang, S.~N., Swartz, D.~A., \& Liu, J.~F. 2009, \apj,
  699, 281

\end{thebibliography}

\end{document}